\documentclass[%
 reprint,
 amsmath,amssymb,
 aps,
]{revtex4-2}

\usepackage{graphicx}
\usepackage{dcolumn}
\usepackage{bm}
\usepackage{xcolor}
\usepackage{changes}
\usepackage[colorlinks=true,linkcolor=black,citecolor=black,urlcolor=black]{hyperref}
\usepackage{booktabs}

\newcommand{\rev}[1]{#1}
\newcommand{\nw}[1]{#1}
\newcommand{\fx}[1]{#1}

\begin{document}


\author{Sk Noor Alam}
\email{noor1989phyalam@gmail.com}
\affiliation{Department of Physics, Aligarh Muslim University\\ Aligarh, Uttar Pradesh-202002, India}
\author{Victor Roy}
\email{victor@niser.ac.in}
\affiliation{School of Physical Sciences, National Institute of Science Education and Research, Bhubaneswar 752050, India.}
\affiliation{Homi Bhabha National Institute, Training School Complex, Anushaktinagar, Mumbai 400094, Maharashtra, India.}

\date{\today}

\title{Collective Flow, Kinetic Freeze-Out, and a Data-Driven Baryon
Chemical Potential in Au+Au Collisions at
$\sqrt{s_{NN}} = 7.7$--$39$ GeV}

\begin{abstract}
We study how transverse and longitudinal collective flow affect kinetic
freeze-out conditions and the net baryon density in 0--5\% central Au+Au
collisions at $\sqrt{s_{NN}} = 7.7$--$39$ GeV, using a covariant statistical
fireball model fitted to STAR identified-hadron transverse-momentum spectra at
three fixed longitudinal velocities $v_z = 0$, $0.2$, and $0.4$.
\rev{Longitudinal flow shifts the extracted temperature upward through a
kinematic $v_z$--$T$ degeneracy rather than by hardening the spectra; imposing
$T \le T_c \approx 155$--$160$ MeV bounds the longitudinal flow to
$v_z \lesssim 0.6$.}
\nw{A simultaneous $\pi^+$, $K^+$, $p$ fit with shared $(T, v_T)$ sharpens the
extraction \fx{by a factor of $3.7$--$5.3$ on $T$ relative to the
single-species proton fit} and gives $T_{\rm kin}\simeq107$--$115$~MeV,
reproducing STAR's independent blast-wave values \fx{to within $0.9\sigma$} and
lying well below the chemical freeze-out temperature.}
\fx{Further including the antiproton spectra, with a normalization shared
between $p$ and $\bar p$, breaks the fugacity--normalization degeneracy and
makes the baryon chemical potential data-driven: the $\bar p/p$ ratio fixes
$\mu_B/T$, the meson spectra fix $T$, and together they give
$\mu_B \simeq 262$--$63$~MeV at kinetic freeze-out, satisfying
$\mu_B/T_{\rm kin} \simeq (0.85$--$0.88)\,\mu_B^{\rm ch}/T_{\rm ch}$: the
$\bar p/p$ ratio is thus largely, but not exactly, set at the chemical surface,
the $12$--$15\%$ deficit being significant at up to $3.2\sigma$.
The resulting kinetic freeze-out trajectory in the $(\rho_B,\varepsilon)$ plane
shows that decoupling occurs in a dilute system, reaching only
$\rho_B \simeq 0.03\,n_0$ at $7.7$~GeV, \fx{a factor of $\simeq25$} below the
chemical-surface compression maximum, while longitudinal flow raises the
inferred \fx{$\rho_B$} by up to a factor of $\sim$4.} These results provide
more precise kinetic freeze-out benchmarks for hydrodynamic modeling and for
future heavy-ion collision programs at FAIR and the RHIC Beam Energy Scan.
\end{abstract}

\maketitle

\section{Introduction}
\label{Section:Introduction}

Mapping the phase diagram of strongly interacting matter, governed by quantum chromodynamics~(QCD), across the space of temperature and chemical potentials (baryon, strangeness, etc.), and locating the critical point, represents one of the principal objectives of heavy-ion collision experiments and theoretical studies. Substantial progress has been made toward achieving this goal~\cite{cbwj-4jfl,sorenson,basar2024qcd,PhysRevLett.114.142301,clarke2025,kanaya2022,AOKI200646,PRD84_071503,rj6r-dmg9}.

Within the QCD phase diagram~\cite{SHAIKH2025122981,endrodi2015,SorensonQCD}, \fx{hadron-yield systematics~\cite{2023137545,9l69-2d7p} show that \emph{chemical} freeze-out, the stage at which inelastic reactions cease and the hadron abundances are fixed, occurs along a smooth curve in the $(T,\mu_B)$ plane~\cite{PhysRevC.74.047901,Andronic:2018Freezeout}. Near $\mu_B\!\approx\!0$, this chemical freeze-out line lies close to the QCD crossover band at $T_c\simeq 155\,\mathrm{MeV}$ established by lattice QCD~\cite{AOKI200646,HotQCD:2014EoS}, thus providing an empirical ``thermometer'' anchoring the phase diagram at small baryon density~\cite{HotQCD:2014EoS}. After chemical freeze-out, the hadron gas continues to expand and cool until \emph{kinetic} freeze-out, when elastic interactions also cease and the momentum spectra are frozen; kinetic freeze-out therefore occurs later and cooler, $T_{\rm kin} \le T_{\rm ch}$, and it is this later surface that transverse-momentum spectra probe.} As $\sqrt{s_{NN}}$ decreases and the baryon chemical potential $\mu_B$ increases, the chemical freeze-out line is expected to pass in the vicinity of any putative QCD critical point. This makes it the natural staging ground for investigating fluctuation observables and non-monotonic trends in Beam Energy Scan (BES) programs~\cite{PhysRevLett.128.202303,PhysRevD.106.094024}. Crucially, Randrup and Cleymans recast the \fx{chemical} freeze-out systematics into \emph{thermodynamic} variables constrained by conservation laws, the net baryon density $\rho_B$ and the energy density $\varepsilon$. They show that $\rho_B$ at freeze-out exhibits a clear maximum at intermediate energies (corresponding to $\mu_B\!\sim\!400$--$500\,\mathrm{MeV}$), pinpointing the region of highest net-baryon compression accessible in heavy-ion collisions and guiding the choice of collision energies at FAIR and RHIC low-energy runs~\cite{PhysRevC.74.047901}. \fx{We extend that construction to the \emph{kinetic} freeze-out surface by incorporating collective flow into the fireball description and re-evaluating the freeze-out mapping in the $(\rho_B,\varepsilon)$ plane across $\sqrt{s_{NN}}$. This allows us to assess how dynamical expansion and the later, cooler decoupling stage modify the net-baryon compression attained at decoupling and, consequently, the proximity of the freeze-out conditions to the conjectured critical region. At the kinetic surface, the net-baryon density is found to be largest at the lowest energy studied, placing the compression maximum at or below $\sqrt{s_{NN}} = 7.7$~GeV, while its magnitude remains strongly sensitive to the assumed longitudinal flow.}

The fireball formed in relativistic heavy-ion collisions is a spatially and 
temporally localized region of hot, dense matter composed of deconfined quarks 
and gluons, known as the Quark-Gluon Plasma (QGP). As the system expands and 
cools, it undergoes a transition to hadronic matter composed of various hadron 
species, well described by a Hadron Resonance Gas (HRG). There is ample evidence 
that both the QGP and HRG phases attain a state of local thermal equilibrium, 
characterized by a temperature $T$ and chemical potential $\mu$. The HRG continues 
to expand until kinetic freeze-out, at which point elastic interactions cease and 
the momentum distributions of the constituent hadrons are fixed; it is these 
distributions that are ultimately recorded in detectors. As a conceptual baseline, 
one may treat the fireball as a static, thermalized source in the laboratory frame, 
from which hadrons are emitted with energies governed by the relativistic Boltzmann 
distribution,
\begin{equation}
f(E) = \lambda \, E \, e^{-E/T},
\label{Eq:eq0}
\end{equation}
where $\lambda$ is a normalization constant and the factor of $E$ arises from the 
momentum-space density of states, $d^3p \propto E\,dE$. In reality, however, the 
fireball undergoes a dynamical evolution driven by internal pressure gradients, 
generating collective flow fields that superimpose a common expansion velocity on 
the underlying thermal motion. As a result, observables such as transverse-momentum 
spectra and particle yields carry the combined imprint of thermal and collective 
dynamics, and neglecting flow biases the freeze-out parameters extracted from thermal fits.

\fx{Concretely, we derive the invariant spectrum emitted by a fireball element having independent longitudinal and transverse flow velocities, and fit it to
STAR identified-hadron $p_T$ spectra measured in central Au+Au collisions at
$\sqrt{s_{NN}}=7.7$--$39$~GeV. The fit is built up in three stages: single-species
fits establish the $v_z$--$T$ kinematic degeneracy; a simultaneous $\pi^+,K^+,p$
fit with shared $(T,v_T)$ fixes the kinetic freeze-out temperature; and the
inclusion of $\bar p$ with a normalization shared with $p$ renders $\mu_B$
data-driven through the $\bar p/p$ ratio. This last step is what distinguishes
the present analysis from a thermal-model fit: both $T$ and $\mu_B$ on the
kinetic surface are determined by the measured spectra alone, with no
chemical-freeze-out systematics used as input, so that the comparison with the
chemical surface is a genuine comparison rather than a consistency check
against an assumed parametrization. The resulting $(T,\mu_B)$ are then
mapped onto $(\rho_B,\varepsilon)$ using a hadron resonance gas, allowing the
kinetic freeze-out trajectory to be compared directly with the chemical one.}

The paper is organized as follows. In Section~\ref{Section:dynamic_fireball}, we present the 
theoretical formulation used to model a dynamically expanding fireball. Section~\ref{Section:Results}
contains our main results, including fits to transverse-momentum spectra across 
a range of center-of-mass energies $\sqrt{s_{NN}}$, from which we extract the 
freeze-out temperature $T$ and baryon chemical potential $\mu_B$, and subsequently 
map these onto the energy density $\varepsilon$ and net baryon density $\rho_B$ as 
functions of $\sqrt{s_{NN}}$. Section~\ref{Section:conclusion} presents our conclusions 
and outlook.

Throughout this paper, we adopt natural units, setting $\hbar = c = k_B = 1$. We choose a mostly negative metric $g_{\mu\nu}=diag(+1,-1,-1,-1)$.
\section{Model of a Dynamic Fireball}
\label{Section:dynamic_fireball}

Before introducing the formulation for particle emission from a dynamically 
expanding thermal fireball, we first review the static fireball framework employed 
in Ref.~\cite{PhysRevC.74.047901}, which serves as the baseline against which our 
approach is contrasted.

In the static fireball scenario, hadrons are assumed to be emitted from a thermal 
source following the Lorentz-invariant differential momentum distribution
\begin{equation}
E\frac{d^3{N}}{d^3{p}} \equiv f(E,p_T,p_z),
\label{Eq:eq1}
\end{equation}
where $E$, $p_T = \sqrt{p_x^2 + p_y^2}$, and $p_z$ are the energy,
transverse momentum, and longitudinal momentum of the particle, respectively. A
more convenient kinematic variable in high-energy collisions is the rapidity $y$, 
defined as
\begin{equation}
y = \tanh^{-1}\!\left(\frac{p_z}{E}\right),
\label{Eq:eq2}
\end{equation}
from which it follows that
\begin{equation}
dy = \frac{dp_z}{E}.
\label{Eq:eq3}
\end{equation}
We further define the transverse mass $m_T = \sqrt{m^2 + p_T^2}$, where $m$ is
the particle mass, and the azimuthal angle $\varphi = \tan^{-1}(p_y/p_x)$. The Lorentz-invariant
phase-space volume element $d^3p = dp_x\,dp_y\,dp_z$ can then be written as
\begin{equation}
d^3{p} = E\,dy\,m_T\,dm_T\,d\varphi,
\label{Eq:eq4}
\end{equation}
with the energy expressed as $E = m_T\cosh y$. Substituting 
Eqs.~\eqref{Eq:eq0} and~\eqref{Eq:eq4} into Eq.~\eqref{Eq:eq1}, the differential 
particle spectrum for the static fireball becomes
\begin{equation}
\frac{d^2N}{m_T^2\,dm_T\,dy} = \lambda \int d\varphi\;\cosh y\; 
e^{-\beta m_T \cosh y},
\label{Eq:eq5}
\end{equation}
where $\beta = 1/T$. The transverse-mass spectrum is obtained by integrating 
Eq.~\eqref{Eq:eq5} over rapidity within the experimentally accepted window 
$[y_\mathrm{low},\, y_\mathrm{up}]$,
\begin{equation}
\frac{1}{m_T}\frac{dN}{dm_T} = \lambda \int d\varphi 
\int_{y_\mathrm{low}}^{y_\mathrm{up}} dy\; m_T \cosh y\; 
e^{-\beta m_T \cosh y}.
\label{Eq:eq6}
\end{equation}

This formulation is used in Ref.~\cite{PhysRevC.74.047901} to evaluate the net 
baryon density and energy density. 

To model the dynamic fireball, we adopt the covariant statistical formulation of 
Ref.~\cite{Touschek}. As noted in the introduction, the fireball undergoes 
dynamical evolution driven by internal pressure gradients, generating collective 
flow fields that superimpose a common expansion velocity on the underlying thermal 
motion. Consider a particle with four-momentum $p^{\mu}$ measured in the rest frame 
of a moving volume element of the fireball. We determine the transformation of the 
intrinsic thermal spectrum defined in the local rest frame of the moving source 
as seen by an arbitrary laboratory observer, expressed entirely in terms of 
Lorentz-covariant quantities. For a comoving volume element $V_0$ at rest in the 
local frame, the Touschek invariant phase-space measure reads~\cite{Touschek}
\begin{equation}
\frac{V_0\,d^3{p}}{(2\pi)^3}\,e^{-E/T} \;\longrightarrow\; 
2\,\frac{V_{\mu}p^{\mu}}{(2\pi)^3}\,d^4{p}\,\delta_{+}(p^2-m^2)\,
e^{-p_{\mu}u^{\mu}/T},
\label{Eq:eq7}
\end{equation}
where $\delta_{+}$ denotes the Dirac delta function restricted to positive energies,
and $V^{\mu} = V_0\,u^{\mu}$ is the four-volume element of the flowing fireball as 
observed in the laboratory frame, with $u^{\mu}$ the local four-velocity field of 
the fireball. A standard calculation yields
\begin{equation}
2\,\delta_{+}(p^2-m^2)\,d^4p = \frac{d^3p}{E} = 
m_T\,dm_T\,dy\,d\varphi = p_T\,dp_T\,dy\,d\varphi.
\label{Eq:eq8}
\end{equation}

Adopting a cylindrical coordinate system consistent with the azimuthal symmetry of 
the collision geometry, the four-momentum of the particle is
\begin{equation}
p^{\mu} = \bigl(m_T\cosh y,\; p_T\cos\varphi,\; p_T\sin\varphi,\; 
m_T\sinh y\bigr).
\label{Eq:eq9}
\end{equation}
\fx{We introduce a transverse rapidity $y_T$ for the particle, its longitudinal
rapidity being the variable $y$ of Eq.~\eqref{Eq:eq2}, via}
\begin{equation}
\cosh y_T = \frac{m_T}{m} \equiv \gamma_T = \frac{1}{\sqrt{1-v_T^2}},
\qquad
\sinh y_T = \frac{p_T}{m} = \gamma_T v_T,
\label{Eq:yT}
\end{equation}
\fx{where $v_T = p_T/m_T$ is the particle's transverse velocity in the frame
reached from the laboratory by a longitudinal boost of rapidity $y$, in which
the particle has no longitudinal momentum; the two relations in
Eq.~\eqref{Eq:yT} are consistent only with this reading, since
$m_T^2 = m^2 + p_T^2$ gives $1-p_T^2/m_T^2 = m^2/m_T^2$. The same convention is
used below for the flow rapidity $y_T^{\prime}$.} Equation~\eqref{Eq:eq9} then becomes
\begin{multline}
\frac{p^{\mu}}{m} = \bigl(\cosh y_T\cosh y,\;\sinh y_T\cos\varphi,\\
\sinh y_T\sin\varphi,\;\cosh y_T\sinh y\bigr),
\label{Eq:eq10}
\end{multline}
Similarly, the four-velocity field of the fireball in cylindrical coordinates is

\begin{multline}
u^{\mu} = \bigl(\cosh y_T^{\prime}\cosh y_{\parallel}^{\prime},\;
\sinh y_T^{\prime}\cos\varphi^{\prime},\\
\sinh y_T^{\prime}\sin\varphi^{\prime},\;
\cosh y_T^{\prime}\sinh y_{\parallel}^{\prime}\bigr),
\label{Eq:eq11}
\end{multline}
with $u^{\mu}u_{\mu} = 1$. Primed quantities ($v^{\prime}$, $y_T^{\prime}$,
$y_{\parallel}^{\prime}$, $\varphi^{\prime}$) refer to the fireball velocity field,
distinguishing it from the velocity of the particle $v$.

Contracting Eqs.~\eqref{Eq:eq10} and~\eqref{Eq:eq11} gives a form symmetric in
the particle and flow rapidities,
\begin{multline}
u_{\mu}p^{\mu} = m\bigl[\cosh y_T\,\cosh y_T^{\prime}\,
\cosh(y - y_{\parallel}^{\prime})\\
-\,\sinh y_T\,\sinh y_T^{\prime}\,\cos(\varphi - \varphi^{\prime})\bigr],
\label{Eq:eq12sym}
\end{multline}
\fx{in which the unprimed factors belong to the particle four-momentum $p^{\mu}$
and the primed ones to the fluid four-velocity $u^{\mu}$. Using the particle
relations $m\cosh y_T = m_T$ and $m\sinh y_T = p_T$ from Eq.~\eqref{Eq:yT} for
the first factor of each pair, and the corresponding flow relation
$\sinh y_T^{\prime} = \cosh y_T^{\prime}\,v_T^{\prime}$ for the second, this
becomes}
\begin{equation}
u_{\mu}p^{\mu} = \cosh y_T^{\prime}\!\left[m_T\cosh(y - y_{\parallel}^{\prime})
- p_T\,v_T^{\prime}\cos(\varphi - \varphi^{\prime})\right],
\label{Eq:eq12}
\end{equation}
\fx{so that the transverse \emph{momentum} $p_T$ is a particle variable while
the transverse \emph{velocity} $v_T^{\prime}$ is a property of the flow; the
particle's own $v_T$ does not appear.}
Substituting Eqs.~\eqref{Eq:eq7} and~\eqref{Eq:eq12} into the general spectrum, 
we obtain the invariant spectrum, which generalises Eq.~\eqref{Eq:eq6},
\begin{multline}
\frac{d^2N}{m_T^2\,dm_T\,dy} = \frac{\lambda\,\gamma_T^{\prime}}{(2\pi)^3}
\int d\psi \\
\left[\cosh(y - y_{\parallel}^{\prime})
- \frac{p_T}{m_T}\,v_T^{\prime}\cos\psi\right] \\
\times\exp\!\left\{-\frac{\gamma_T^{\prime}}{T}
\left[m_T\cosh(y-y_{\parallel}^{\prime})
- p_T v_T^{\prime}\cos\psi\right]\right\},
\label{Eq:eq13}
\end{multline}
where $\psi = \varphi - \varphi^{\prime}$ and 
$\gamma_T^{\prime} = \cosh y_T^{\prime} = 1/\sqrt{1 - v_T^{\prime\,2}}$.

\fx{The integrand of Eq.~\eqref{Eq:eq13} depends on the two azimuths only
through their difference $\psi = \varphi - \varphi^{\prime}$. For a single
volume element, we may therefore choose the coordinate system so that
$\varphi^{\prime} = 0$, giving $\psi = \varphi$; this is a relabeling of the
azimuthal origin and involves no physical assumption. The transverse flow of
the fireball as a whole is of course directed radially outward, so
$\varphi^{\prime}$ varies from element to element and cannot be set to zero
globally. What makes Eq.~\eqref{Eq:eq15} nonetheless applicable to the whole
source is azimuthal symmetry: for the central collisions considered here, the
elements are distributed uniformly in $\varphi^{\prime}$ and carry a radial
flow of common magnitude $v_T^{\prime}$, so integrating over $\varphi^{\prime}$
contributes only an overall factor $2\pi$, absorbed into $\mathcal{N}$ in
Eq.~\eqref{Eq:lambdadef}, and leaves the $\psi$-dependence unchanged. Each
element then contributes identically to the azimuthally integrated spectrum,
and $v_T^{\prime}$ is to be read as a single effective radial flow velocity
rather than a local one; we do not resolve a radial velocity profile, and
azimuthal anisotropy is absent by construction.} Since $\cos\psi$ is even, we may restrict
the integration to $\psi \in [0, \pi]$ and multiply by a factor of 2. The
$\psi$-integral is then identified with the modified Bessel function of the
first kind,
\begin{equation}
I_n(a) = \frac{1}{\pi}\int_{0}^{\pi} e^{a\cos\psi}\cos(n\psi)\,d\psi,
\label{Eq:eq14}
\end{equation}
which we evaluate numerically. Performing the azimuthal integration in 
Eq.~\eqref{Eq:eq13} yields
\begin{multline}
\frac{d^2N}{m_T^2\,dm_T\,dy} = \frac{\lambda\,\gamma_T^{\prime}}{(2\pi)^2}
\left[\cosh(y - y_{\parallel}^{\prime})\,I_0(a) 
- \frac{p_T}{m_T}v_T^{\prime}\,I_1(a)\right] \\
\times\exp\!\left\{-\frac{\gamma_T^{\prime}\,m_T\cosh(y-y_{\parallel}^{\prime})}{T}
\right\},
\label{Eq:eq15}
\end{multline}
where $a = \gamma_T^{\prime}\,p_T\,v_T^{\prime}/T$.

Equation~\eqref{Eq:eq15} gives the invariant particle spectrum emitted from a 
volume element of the dynamic fireball with transverse and longitudinal flow 
velocities $v_T^{\prime}$ and $v_{\parallel}^{\prime}$, as observed in the 
laboratory frame. \fx{The prefactor $\lambda$ is an overall normalization,
\begin{equation}
\lambda \;=\; \mathcal{N}\,e^{\mu_B B/T},
\label{Eq:lambdadef}
\end{equation}
in which $\mathcal{N}$ collects the comoving volume and all $p_T$-independent
experimental factors, and $e^{\mu_B B/T}$ is the fugacity associated with the
conserved baryon number $B$ of the species ($B=0$ for mesons). Only the product
enters the spectrum; we return to the consequences of this in
Sec.~\ref{Section:Results}.} Since the fireball
is characterised by four independent parameters (the temperature $T$, the baryon 
chemical potential $\mu_B$, the transverse flow velocity $v_T^{\prime}$, and the 
longitudinal flow velocity $v_{\parallel}^{\prime}$), we treat all four as free 
parameters and determine them by fitting Eq.~\eqref{Eq:eq15} to experimental
transverse-momentum spectra.

\fx{We note that Eq.~\eqref{Eq:eq7} has the structure of a Cooper--Frye
emission integral~\cite{CooperFrye1974} for a freeze-out element whose surface
normal is proportional to $u^\mu$, and Eq.~\eqref{Eq:eq15} may correspondingly
be viewed as a covariant, single-surface-velocity variant of the blast-wave
spectrum of Schnedermann, Sollfrank, and Heinz~\cite{Schnedermann1993}. Unlike
the standard blast wave, it retains an explicit longitudinal flow velocity and
does not average over a transverse velocity profile; the fitted $v_T'$ should
therefore be compared with a surface, rather than profile-averaged, transverse
velocity. We also note that the parametrization of Eq.~\eqref{Eq:eq11} is a longitudinal
boost of rapidity $y_{\parallel}'$ applied to a purely transverse flow: boosting
$u^{\mu}$ by $-y_{\parallel}'$ along $z$ gives
$(\cosh y_T',\,\sinh y_T'\cos\varphi',\,\sinh y_T'\sin\varphi',\,0)$. Hence
$v_{\parallel}' = \tanh y_{\parallel}'$ is the longitudinal flow velocity in the
laboratory frame, whereas $v_T' = \tanh y_T'$ is the transverse flow velocity
measured in the longitudinally comoving frame, in which the flow has no
longitudinal component. The laboratory-frame transverse flow velocity is
correspondingly $\tanh y_T'/\cosh y_{\parallel}'$, smaller by one factor of
$\cosh y_{\parallel}'$; the identification
$\gamma_T' = \cosh y_T' = 1/\sqrt{1-v_T'^{\,2}}$ refers to the comoving
transverse velocity.}

\fx{A remark on notation. Throughout this section primes distinguish the
fireball velocity field from the particle kinematics, the latter entering only
through Eqs.~\eqref{Eq:eq9}--\eqref{Eq:eq10}. Since the particle transverse
velocity $v_T$ of Eq.~\eqref{Eq:yT} plays no further role once the spectrum
Eq.~\eqref{Eq:eq15} has been obtained (the particle kinematics entering
thereafter only through the measured $p_T$, $m_T$ and $y$), we drop the
primes from here on and
write $v_T \equiv v_T'$ and $v_z \equiv v_{\parallel}'$ for the transverse and
longitudinal \emph{flow} velocities of the fireball. All velocities quoted in
Secs.~\ref{Section:Results}--\ref{Section:conclusion}, and in every figure and
table, are fireball flow velocities in this sense.}
\subsection{Calculation of Energy Density and Net Baryon Density}

\fx{The spectrum of Eq.~\eqref{Eq:eq15} is controlled by the freeze-out
temperature $T$ and, through the fugacity of Eq.~\eqref{Eq:lambdadef}, by the
baryon chemical potential $\mu_B$; the fits that extract these from the
measured spectra are presented in Sec.~\ref{Section:Results}. Since the
freeze-out systematics of Ref.~\cite{PhysRevC.74.047901} are formulated not in
$(T,\mu_B)$ but in the thermodynamic variables fixed by conservation laws, a
mapping from the former to the latter is required. We set out that mapping
here, before turning to the data.} Specifically, we employ the grand-canonical
partition function $Z_i$ for each hadron species $i$ to express the energy
density $\varepsilon$ and the net baryon density $\rho_B$ of a hadron resonance
gas at a given $(T,\mu_B)$. The total partition function factorizes over all species as $Z = \prod_{i} Z_i$, where $Z_i$ depends on the temperature $T$, the volume element $V$, and the baryon, electric charge, and strangeness chemical potentials $\mu_B$, $\mu_Q$, and $\mu_S$, respectively,
\begin{equation}
\ln Z_i(T, V, \{\mu\}) = \pm \frac{V T\, g_i m_i^2}{2\pi^2} \sum_{n=1}^{\infty} \frac{(\pm\lambda_i)^n}{n^2}\, K_2\!\left(\frac{n m_i}{T}\right),
\label{Eq:partion0}
\end{equation}
where the upper (lower) sign applies to bosons (fermions), $m_i$ is the mass of species $i$, $g_i = 2J_i + 1$ is the spin degeneracy, and $\lambda_i$ is the fugacity
\begin{equation}
\lambda_i(T, \{\mu\}) = \exp\!\left(\frac{\mu_B B_i + \mu_Q Q_i + \mu_S S_i}{T}\right),
\label{Eq:fugacity}
\end{equation}
with $B_i$, $Q_i$, and $S_i$ denoting the baryon number, electric charge, and strangeness of species $i$, respectively. \fx{The subscripted $\lambda_i$ of Eq.~\eqref{Eq:fugacity} is the species fugacity in the conventional hadron-resonance-gas notation, and is not to be confused with the fitted spectral normalization $\lambda$ of Eq.~\eqref{Eq:lambdadef}.}

The number density of species $i$ follows from the standard thermodynamic relation
\begin{equation}
n_i(T, \{\mu\}) = \frac{T}{V}\left(\frac{\partial \ln Z_i}{\partial \mu}\right)_{T,V},
\label{Eq:numberDen1}
\end{equation}
which, upon substitution of Eq.~\eqref{Eq:partion0}, yields
\begin{align}
n_i(T, \{\mu\}) &= \pm\frac{g_i T^3}{2\pi^2} \sum_{n=1}^{\infty} \frac{(\pm\lambda_i)^n}{n^3} \left(\frac{n m_i}{T}\right)^2 K_2\!\left(\frac{n m_i}{T}\right) \notag \\
&\approx \frac{g_i \lambda_i}{2\pi^2}\, m_i^2\, T\, K_2\!\left(\frac{m_i}{T}\right) \pm \ldots
\label{Eq:numerDen2}
\end{align}
\rev{For the thermodynamic densities reported below, we retain the full
Bose--Einstein/Fermi--Dirac series (summed to convergence). The leading
($n=1$) term already provides an excellent approximation for the heavy
baryons that dominate the net-baryon density, whereas the higher terms give a
non-negligible contribution for light mesons (pions) and are kept for the
energy density.} The net baryon density is then obtained by summing the number densities weighted by the baryon number $B_i$,
\begin{equation}
\rho_B(T, \{\mu\}) = T\frac{\partial \ln Z}{\partial \mu_B} = \sum_{i} B_i\, n_i(T, \{\mu\}),
\label{Eq:netBaryon}
\end{equation}
where the sum runs over all hadron species.

The energy density of species $i$ is given by $\varepsilon_i(T, \{\mu\}) = -\partial \ln Z_i / \partial\beta$, where $\beta = 1/T$. Evaluating this derivative yields
\begin{multline}
\varepsilon_i(T, \{\mu\}) = \pm\frac{g_i T^4}{2\pi^2} \sum_{n=1}^{\infty} \frac{(\pm\lambda_i)^n}{n^4} \left(\frac{n m_i}{T}\right)^2 \\
\times \left[\rev{3}\,K_2\!\left(\frac{n m_i}{T}\right) + \frac{n m_i}{T}\, K_1\!\left(\frac{n m_i}{T}\right)\right].
\label{Eq:energyDen1}
\end{multline}
\rev{The coefficient $3$ multiplying $K_2$ reflects the mean thermal energy per
particle, $\langle E\rangle/N = 3T + m\,K_1(m/T)/K_2(m/T)$; equivalently
$\varepsilon_i = n_i\,[\,3T + m_i K_1(m_i/T)/K_2(m_i/T)\,]$.} Keeping the leading term, the energy density simplifies to
\begin{equation}
\varepsilon_i(T, \{\mu\}) \approx \frac{g_i \lambda_i}{2\pi^2}\, m_i^2\, T^2 \left[\rev{3}\,K_2\!\left(\frac{m_i}{T}\right) + \frac{m_i}{T}\, K_1\!\left(\frac{m_i}{T}\right)\right].
\label{Eq:energyDen2}
\end{equation}
The total energy density is $\varepsilon = \sum_i \varepsilon_i$. \fx{Equations~\eqref{Eq:numerDen2}, \eqref{Eq:netBaryon}, and \eqref{Eq:energyDen2} thus determine $\rho_B$ and $\varepsilon$ for any given $(T,\mu_B)$. In Sec.~\ref{Section:Results} they are evaluated at the freeze-out parameters obtained from the fits, yielding the kinetic freeze-out trajectory in the $(\rho_B,\varepsilon)$ plane across the full range of collision energies considered.}

\section{Results and Discussion}
\label{Section:Results}

In this section, we present and discuss our main results. We begin by examining the best fits of Eq.~\eqref{Eq:eq15} to the experimentally measured transverse-momentum ($p_T$) spectra of identified hadrons, specifically, positive pions and protons in 0--5\% central Au+Au collisions measured by the STAR Collaboration at $\sqrt{s_{NN}} = 7.7$--$39$ GeV~\cite{PhysRevC.96.044904}. \fx{These fits determine the temperature $T$, the transverse flow velocity $v_T$ and the normalization $\lambda$ for three fixed values of the longitudinal velocity $v_z = 0$, $0.2$, and $0.4$; the baryon chemical potential is not accessible at this stage, for the reason given below, and is extracted from the data only in Sec.~\ref{Section:mubextraction}.} We then compute particle yields by integrating the fitted spectra over the rapidity range $|y| < 0.1$ and the transverse momentum range $0.2 < p_T < 2.0$ GeV/$c$, and examine their dependence on $v_T$ at each fixed $v_z$ in order to quantify the effect of collective flow on particle abundances.

\rev{For a single hadron species, $\mu_B$ enters the spectrum,
Eq.~\eqref{Eq:eq15}, only through the fugacity $e^{\mu_B B/T}$, which is
$p_T$-independent and therefore multiplies a shape function of $(T,v_T,v_z)$
without distorting it. \fx{By Eq.~\eqref{Eq:lambdadef} the fugacity is not
separable from the volume factor $\mathcal{N}$: the data constrain only their
product $\lambda = \mathcal{N}e^{\mu_B B/T}$, and since $\mathcal{N}$ is itself
unknown, $\mu_B$ and the normalization are exactly degenerate.} The
mid-rapidity proton spectrum fixes the spectral \emph{shape}, and hence $T$
and $v_T$, but carries no independent information on $\mu_B$. We verified this
explicitly: scanning $\mu_B$ over a wide range at fixed $(T,v_T)$ and
re-optimizing the normalization leaves the $\chi^2$ unchanged. We therefore
do not treat $\mu_B$ as a free parameter. Instead we hold it fixed at each
energy while fitting only the temperature
$T$, the transverse flow velocity $v_T$, and the normalization $\lambda$ to
the measured proton $p_T$ spectra.}

\fx{A value for the fixed $\mu_B$ is nonetheless required in order to report
one, and for this we take the chemical freeze-out systematics of
Cleymans~et~al.~\cite{PhysRevC.73.034905},}
\begin{equation}
\mu_B(\sqrt{s_{NN}}) = \frac{d}{1 + e\,\sqrt{s_{NN}}},
\label{eq:mubS}
\end{equation}
with $d = 1.308 \pm 0.028$ GeV and $e = 0.273 \pm 0.008$ GeV$^{-1}$, together
with the associated freeze-out temperature curve
\begin{equation}
T(\mu_B) = a - b\,\mu_B^2 - c\,\mu_B^4,
\label{eq:mubT}
\end{equation}
with $a = 0.166 \pm 0.002$ GeV, $b = 0.139 \pm 0.016$ GeV$^{-1}$, and $c = 0.053 \pm 0.021$ GeV$^{-3}$. These parametrizations were obtained from thermal model fits to particle yields at \emph{chemical} freeze-out, prior to the BES program.

\fx{Their use in a \emph{kinetic} freeze-out analysis therefore requires
comment. Equation~\eqref{eq:mubS} is not a constraint on the fit but a
placeholder for a quantity the single-species data do not determine, and the
degeneracy established above guarantees that its value cannot propagate into
any fitted quantity. We have verified this directly: rescaling the input
$\mu_B$ over the range $0.25$--$4$ times Eq.~\eqref{eq:mubS} shifts the fitted
$T$ by less than $3\times10^{-5}$~MeV and $v_T$ by less than $10^{-7}$,
leaving $\chi^2$ unchanged to one part in $10^{11}$, at every energy, at each
of the three $v_z$ values, and for both the single-species and the
three-species fits; the normalization absorbs the change exactly. The
temperatures and flow velocities of this section are thus independent of the
chemical-freeze-out input, and the choice affects only the $\mu_B$ column
echoed back in Table~\ref{table:fitparams} and any density
evaluated from it. Equation~\eqref{eq:mubT} is likewise never used as a fit
function; it enters only as the reference chemical-freeze-out line against
which the fitted kinetic temperatures are compared, which is precisely the role
a chemical-freeze-out parametrization should play here. Neither equation is
used in the determination of $\mu_B$ in Sec.~\ref{Section:mubextraction}, where
it serves at most as a starting value for the minimizer, and the
$(\rho_B,\varepsilon)$ trajectory reported there is built entirely on the
data-driven $\mu_B$.}

\rev{A genuinely data-driven $\mu_B$ would
require a simultaneous multi-species fit (e.g.\ $p$, $\bar p$, $\pi^{\pm}$,
$K^{\pm}$), in which the relative normalizations break the degeneracy; \fx{this
is carried out in Sec.~\ref{Section:mubextraction}, where the inclusion of the
antiproton spectra determines $\mu_B$ directly from the data}. The resulting
energy-dependent $T$ and (input) $\mu_B$, together with the full set of fitted
parameters and their uncertainties, are reported in
Table~\ref{table:fitparams}. The fitted
kinetic freeze-out temperatures lie somewhat \emph{below} the Cleymans
chemical-freeze-out curve, as expected physically: kinetic freeze-out occurs
later and cooler than chemical freeze-out, and the transverse flow
($v_T \simeq 0.47$--$0.55$) accounts for part of the spectral hardening that a
purely thermal fit would otherwise absorb into a higher $T$.}

\subsection{Fitting the Transverse Momentum Spectra}

\begin{figure}[htb]
\centering
\includegraphics[width=0.47\textwidth]{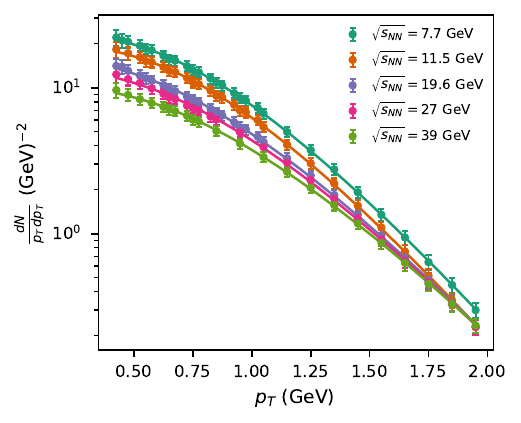}
\caption{(Color online) Midrapidity ($|y| <0.1$) transverse momentum spectra for positive protons in Au+Au collisions at $\sqrt{s_{NN}} = $ 7.7 to 39 GeV for 0-5$\%$ centrality with longitudinal velocity $v_z$ = 0.0 of the dynamic fireball. \fx{Symbols with error bars are the STAR data~\cite{PhysRevC.96.044904}; the solid curves are the corresponding best fits of Eq.~\eqref{Eq:eq15}, each plotted in the same color as the data set it describes.}}
\label{fig:protonpt}
\end{figure}

\begin{figure}[htb]
\centering
\includegraphics[width=0.47\textwidth]{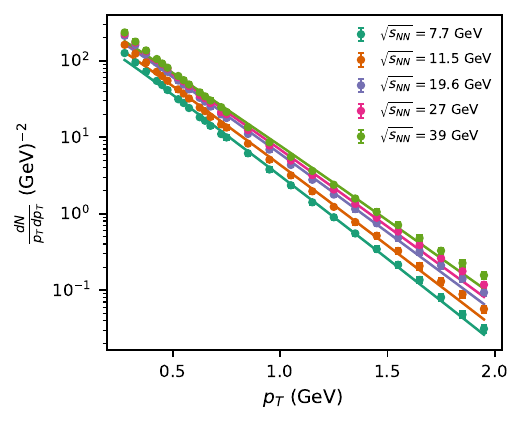}
\caption{(Color online) Midrapidity ($|y| <0.1$) transverse momentum spectra for positive pions in Au+Au collisions at $\sqrt{s_{NN}} = $ 7.7 to 39 GeV for 0-5$\%$ centrality with longitudinal velocity $v_z$ = 0.0 of the dynamic fireball. \fx{Symbols with error bars are the STAR data~\cite{PhysRevC.96.044904}; the solid curves are the corresponding best fits of Eq.~\eqref{Eq:eq15}, each plotted in the same color as the data set it describes.} \rev{The single-freeze-out Boltzmann spectrum reproduces the proton slope closely but describes the pions only approximately at low $p_T$, where resonance feed-down (not included here) is significant.}}
\label{fig:pionpt}
\end{figure}

Figures~\ref{fig:protonpt} and~\ref{fig:pionpt} show the transverse momentum spectra of protons and positive pions, respectively, fitted to STAR measurements in 0--5\% central Au+Au collisions at $\sqrt{s_{NN}} = 7.7$--$39$ GeV within the rapidity range $|y| < 0.1$. \fx{Symbols with error bars are the STAR data and the solid curves are the corresponding best fits, each drawn in the same color as the data set it describes. The curves are obtained by minimizing $\chi^2$ with respect to $(T, v_T, \lambda)$ and then evaluating Eq.~\eqref{Eq:eq15} at the resulting best-fit parameters on a continuous $p_T$ grid.}

The fits describe the experimental data well across the full range of collision energies considered, as confirmed by the \fx{$\chi^2/\mathrm{NDF}$ values for the proton fits reported in Table~\ref{table:fitparams} for all three longitudinal velocities}. \rev{With $\mu_B$ fixed to the Cleymans systematics, the free parameters are $(T,v_T,\lambda)$, so that the number of degrees of freedom equals the number of $p_T$ points minus three.} \fx{The fit quality is essentially independent of $v_z$, degrading by only $2$--$6\%$ in $\chi^2$ between $v_z = 0$ and $v_z = 0.4$ at every energy. This is a direct expression of the fact, discussed further in Sec.~\ref{Section:corr}, that mid-rapidity spectra carry almost no information on the longitudinal flow: the data cannot distinguish between the three $v_z$ values, and $v_z$ is accordingly treated as a fixed external parameter rather than fitted.}
\fx{The weights $\sigma_i$ entering the $\chi^2$ are the total uncertainties
published with the STAR invariant yields (the quadrature sum of the
statistical and systematic components), rescaled by the same
azimuthal/rapidity factor as the yields themselves. Over the $p_T$ windows
fitted here these are strongly dominated by the systematic component: the
median ratio of statistical to systematic uncertainty is $0.03$--$0.08$
depending on species, so that $\sigma_i$ is in effect the systematic error,
amounting to $6$--$12\%$ of the yield. This has a direct consequence for the
reduced $\chi^2$. Systematic uncertainties of this type (normalization,
tracking and particle-identification efficiencies) are largely \emph{correlated}
from one $p_T$ bin to the next, whereas the diagonal $\chi^2$ used here treats
them as independent. The reduced $\chi^2/\mathrm{NDF}$ is therefore expected to
fall well below unity, and indeed does so for the single-species fits of
Table~\ref{table:fitparams} ($\chi^2/\mathrm{NDF} \simeq 0.03$--$0.08$); this
reflects the error treatment rather than an anomalously accurate description of
the data. The parameter uncertainties quoted throughout are the corresponding
diagonal covariance errors, with no rescaling by $\chi^2/\mathrm{NDF}$ applied
anywhere in this work. They are conservative in magnitude but do not carry a
strict frequentist interpretation, and they should be read together with the
strong $T$--$v_T$ correlation discussed in Sec.~\ref{Section:corr} rather than
as independent one-parameter errors.}


\rev{The fitted transverse flow is stable across the whole energy range,
$v_T \simeq 0.47$--$0.55$, rising slightly with $v_z$, while the temperature
increases monotonically with $v_z$ at fixed energy; the associated parameter
correlations are discussed in Sec.~\ref{Section:corr}.}


\fx{In evaluating Eq.~\eqref{Eq:eq15}, we integrate over the same rapidity
window as the data, $|y| < 0.1$, rather than evaluating the spectrum at
$y = 0$.}



\begin{table}[h!]
\centering
\renewcommand{\arraystretch}{1.15}
\begin{tabular}{||c c c c c c c||}
 \hline
 $\sqrt{s_{NN}}$ & \fx{NDF} & $v_z$ & $T$ & $v_T$ & $\mu_B$ & $\chi^2/\mathrm{NDF}$ \\ [0.5ex]
 (GeV) & & & (GeV) & & (GeV) & \\
 \hline\hline
    &         & 0.0 & \rev{$0.121\pm0.007$} & \rev{$0.483\pm0.019$} & \rev{0.422} & \fx{0.031} \\
 7.7 & \fx{26} & 0.2 & \rev{$0.124\pm0.007$} & \rev{$0.493\pm0.019$} & \rev{0.422} & \fx{0.031} \\
    &         & 0.4 & \rev{$0.134\pm0.007$} & \rev{$0.525\pm0.020$} & \rev{0.422} & \fx{0.032} \\
 \hline
    &         & 0.0 & \rev{$0.119\pm0.007$} & \rev{$0.483\pm0.019$} & \rev{0.316} & \fx{0.040} \\
 11.5 & \fx{25} & 0.2 & \rev{$0.122\pm0.007$} & \rev{$0.493\pm0.020$} & \rev{0.316} & \fx{0.041} \\
    &         & 0.4 & \rev{$0.132\pm0.007$} & \rev{$0.526\pm0.021$} & \rev{0.316} & \fx{0.042} \\
 \hline
    &         & 0.0 & \rev{$0.133\pm0.008$} & \rev{$0.471\pm0.022$} & \rev{0.206} & \fx{0.051} \\
 19.6 & \fx{26} & 0.2 & \rev{$0.136\pm0.009$} & \rev{$0.480\pm0.022$} & \rev{0.206} & \fx{0.052} \\
    &         & 0.4 & \rev{$0.148\pm0.009$} & \rev{$0.512\pm0.023$} & \rev{0.206} & \fx{0.053} \\
 \hline
    &         & 0.0 & \rev{$0.134\pm0.010$} & \rev{$0.482\pm0.025$} & \rev{0.156} & \fx{0.075} \\
 27 & \fx{20} & 0.2 & \rev{$0.137\pm0.010$} & \rev{$0.492\pm0.026$} & \rev{0.156} & \fx{0.076} \\
    &         & 0.4 & \rev{$0.149\pm0.011$} & \rev{$0.525\pm0.027$} & \rev{0.156} & \fx{0.079} \\
 \hline
    &         & 0.0 & \rev{$0.136\pm0.010$} & \rev{$0.502\pm0.024$} & \rev{0.112} & \fx{0.028} \\
 39 & \fx{19} & 0.2 & \rev{$0.140\pm0.010$} & \rev{$0.512\pm0.024$} & \rev{0.112} & \fx{0.028} \\
    &         & 0.4 & \rev{$0.152\pm0.011$} & \rev{$0.546\pm0.025$} & \rev{0.112} & \fx{0.030} \\ [1ex]
 \hline
\end{tabular}
\caption{\rev{Fitted freeze-out temperature $T$ and transverse flow velocity
$v_T$ ($1\sigma$ uncertainties from the fit covariance) obtained from the
proton $p_T$ spectra using Eq.~\eqref{Eq:eq15}, at each collision energy and
each fixed longitudinal velocity $v_z$. The baryon chemical potential $\mu_B$
is an input fixed to Eq.~\eqref{eq:mubS}, hence independent of $v_z$; the
normalization $\lambda$ is a nuisance parameter and is not tabulated.
Parameter correlations are given in Table~\ref{table:corr}.} \fx{NDF is
$v_z$-independent and is quoted once per energy.}}
\label{table:fitparams}
\end{table}

\begin{figure*}[htb]
\centering

\includegraphics[width=0.60\textwidth]{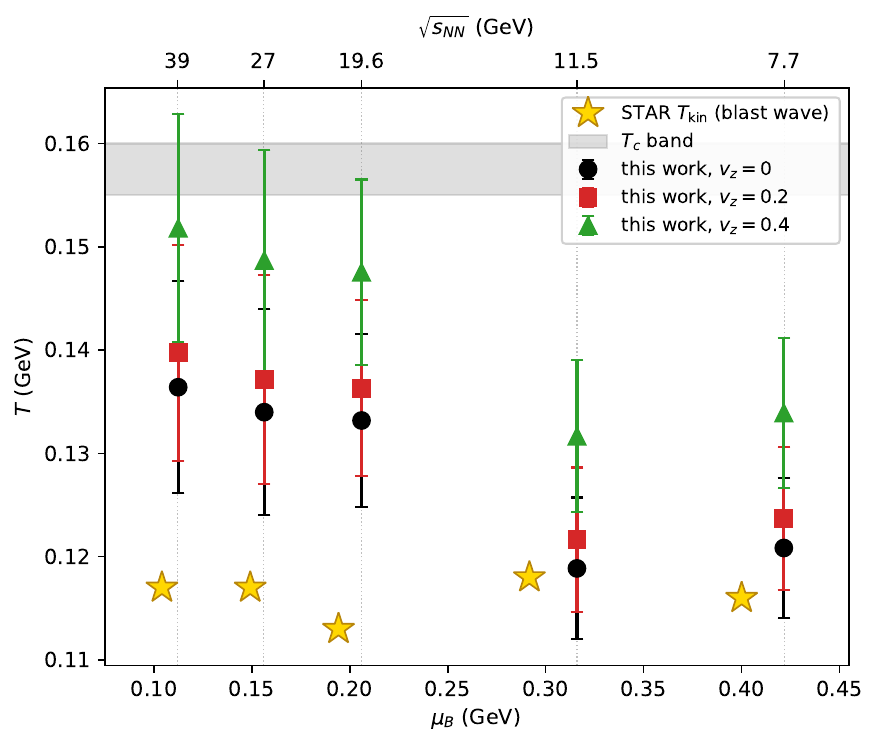}
\caption{(Color online)  \rev{Fitted kinetic freeze-out temperature $T$ versus the (input) baryon chemical potential $\mu_B$} for three longitudinal velocities $v_z =0$ (black circles), $v_z =0.2$ (red squares), and $v_z =0.4$ (green triangles). \rev{Error bars are the $1\sigma$ fit uncertainties on $T$. The shaded band marks the lattice QCD crossover temperature $T_c \approx 155$--$160$ MeV; all fitted points for $v_z \leq 0.4$ lie at or below it.} \nw{Gold stars are STAR's published kinetic freeze-out temperatures $T_{\rm kin}$ from an independent blast-wave analysis, plotted at STAR's measured chemical $\mu_B$~\cite{PhysRevC.96.044904}; they fall just below our $v_z=0$ points, consistent with a kinetic freeze-out extraction.} \fx{The upper axis gives the collision energy $\sqrt{s_{NN}}$ associated with each $\mu_B$ through Eq.~\eqref{eq:mubS}, and the dotted vertical guides mark those five values, so that each column of points can be read directly as one energy. The small horizontal offset ($\lesssim 10\%$) between STAR's measured $\mu_B$ and the Cleymans input used for our points is visible at the lowest energies as a displacement of the stars from the guides.}}
\label{fig:t-mub}
\end{figure*}

\begin{figure*}[htb]
\centering
\includegraphics[width=0.60\textwidth]{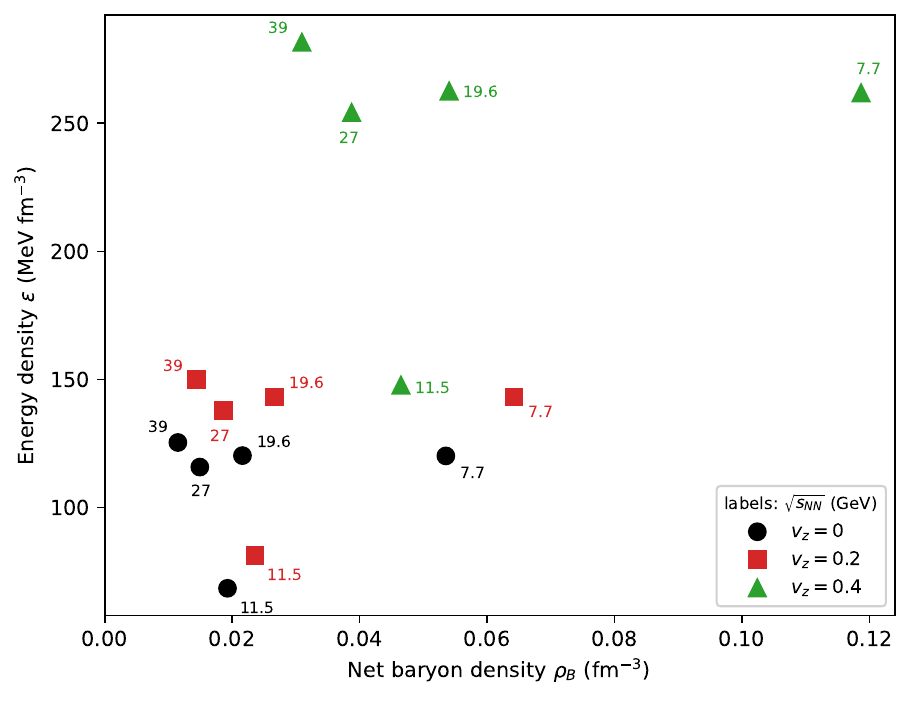}
\caption{(Color online)  \fx{Kinetic freeze-out points in the $(\rho_{B},\varepsilon)$ plane, obtained by evaluating the statistical model at the fitted kinetic freeze-out temperature $T$ and the input baryon chemical potential $\mu_B$, for the different STAR energies and longitudinal fireball velocities. Each point is labelled by its collision energy $\sqrt{s_{NN}}$ in GeV. The low-lying point in each $v_z$ series is $11.5$~GeV, where the fitted $T$ is at its minimum (Table~\ref{table:fitparams}) while $\mu_B$ has already dropped well below its $7.7$~GeV value; the excursion is $\lesssim1\sigma$ once the uncertainty on $T$ is propagated, and it is absent from the fully data-driven evaluation of Fig.~\ref{fig:v4rhoeps}.} }
\label{fig:t-rhoE}
\end{figure*}



%
We now use the extracted parameters to compute the proton yield by integrating
Eq.~\eqref{Eq:eq15} over rapidity $|y| < 0.1$ and transverse momentum
$0.425 \leq p_T \leq 1.95$ GeV \fx{(the range adopted directly from the
STAR data)}, for 0--5\% central Au+Au collisions at $\sqrt{s_{NN}} = 7.7$--$39$
GeV. The resulting proton yields are
tabulated in Table~\ref{table:yield-proton}.

\begin{table}[h!]
\centering
\begin{tabular}{||c c ||}
 \hline
 $\sqrt{s_{NN}}$ (GeV) & $N_p$ \\ [0.5ex]
 \hline\hline
7.7    &\rev{8.503}\\
11.5   &\rev{6.991}\\
19.6   &\rev{5.468}\\
27     &\rev{4.924}\\
39     &\rev{4.202}\\
\hline
\end{tabular}
\caption{\fx{Integrated proton yields at $\sqrt{s_{NN}}$ = 7.7--39 GeV in
0--5\% central Au+Au collisions. $N_p$ is the dimensionless per-event proton
number obtained by integrating Eq.~\eqref{Eq:eq15} over $|y| < 0.1$ and
$0.425 \leq p_T \leq 1.95$~GeV at the best-fit parameters of
Table~\ref{table:fitparams}; it is not $dN/dy$. Values are quoted at
$v_z = 0$; they change by less than $0.02\%$ at $v_z = 0.2$ and $0.4$, so a
single column suffices (see text). The $\mu_B$ entering the fugacity is the
Cleymans input of Eq.~\eqref{eq:mubS}, not the data-driven value of
Sec.~\ref{Section:mubextraction}.}}
\label{table:yield-proton}
\end{table}

Two notable features emerge from these results. First, the proton yield 
decreases monotonically with increasing collision energy, directly reflecting 
the energy dependence of \textit{baryon stopping}. At lower $\sqrt{s_{NN}}$, 
the colliding Au nuclei are relatively opaque to one another and a significant 
fraction of the incoming nucleons lose most of their longitudinal momentum 
through multiple inelastic scatterings and come to rest near mid-rapidity, 
producing a large net-baryon density in the central region. This is 
quantitatively captured by the large values of $\mu_B$ extracted from our 
fits at lower energies. As $\sqrt{s_{NN}}$ increases, the nuclei become 
progressively more transparent, and the incoming baryons largely retain their 
longitudinal momentum, being carried to forward and backward rapidities 
($|y| \gg 0$) while leaving a nearly baryon-free region at mid-rapidity. 
Consequently, $\mu_B \rightarrow 0$ at high energies, and the proton yield 
at mid-rapidity diminishes accordingly. The increasing availability of 
center-of-mass energy also drives copious production of quark-antiquark pairs, 
manifesting as a sharp rise in pion yields with $\sqrt{s_{NN}}$, as clearly 
seen in Fig.~\ref{fig:pionpt}. Thus, as the collision energy increases, the 
character of particle production at mid-rapidity transitions from 
baryon-dominated to meson-dominated, a signature of the approach toward 
net-baryon transparency.

Second, the proton yield is found to be essentially independent of the
longitudinal fireball velocity $v_z$. \fx{The cancellation is striking: going
from $v_z = 0$ to $v_z = 0.4$ raises the fitted temperature by about $11\%$ at
every energy, yet changes the integrated yield by less than $0.02\%$. The
normalization $\lambda$ absorbs the change almost exactly, in the same way that
it absorbs a rescaling of the input $\mu_B$. This is why
Table~\ref{table:yield-proton} quotes a single value per energy.} This can be
understood as follows. The
net-baryon content at mid-rapidity is established during the very early, 
pre-equilibrium stage of the collision through baryon stopping, well before 
any collective flow develops. Collective longitudinal flow, characterized by 
$v_z$, acts as a subsequent rigid rapidity boost of the expanding fireball. 
For the narrow rapidity window $|y| < 0.1$ considered here, such a boost 
shifts the rapidity distribution by $\Delta y \sim \tanh^{-1}(v_z)$, but 
since the net-baryon rapidity distribution is broad and approximately flat 
near mid-rapidity at these energies, the integrated yield within this small 
window remains largely unaffected. In other words, the conserved baryon charge 
within $|y| < 0.1$ is insensitive to the collective longitudinal motion of 
the fireball, as long as $v_z$ remains moderate, consistent with our chosen 
values of $v_z \leq 0.4$.

 \fx{Each freeze-out point is formed by pairing the fitted kinetic temperature
$T$ of Table~\ref{table:fitparams} with the chemical-freeze-out baryon chemical
potential $\mu_B(\sqrt{s_{NN}})$ of Eq.~\eqref{eq:mubS}, which depends only on
the collision energy and is therefore common to all three $v_z$.} This is done
at each collision energy for 0--5\%
central Au+Au collisions and for each of the three fixed longitudinal
velocities $v_z = 0.0$, $0.2$, and $0.4$. The resulting $T$--$\mu_B$
freeze-out points are displayed in Fig.~\ref{fig:t-mub} \rev{and listed in
Table~\ref{table:fitparams}}.

Several trends are immediately apparent. \fx{The input $\mu_B$ of
Eq.~\eqref{eq:mubS} decreases monotonically with $\sqrt{s_{NN}}$ by
construction}, consistent with the picture of increasing nuclear
transparency and diminishing net-baryon stopping at mid-rapidity discussed
above. \fx{The fitted $T$ rises overall, spanning $119$--$136$~MeV at
$v_z = 0$, reflecting the larger energy density deposited in the collision
zone at higher energies; the one exception is a $2$~MeV dip at $11.5$~GeV,
well inside the $\pm 7$~MeV fit uncertainties of
Table~\ref{table:fitparams} ($0.2\sigma$) and discussed below in connection
with $\rho_B$.}
Together, these trends trace a freeze-out curve in the $T$--$\mu_B$ plane 
that moves from the high-$\mu_B$, low-$T$ region (low energies) toward the 
low-$\mu_B$, high-$T$ region (high energies)\fx{, a trend that qualitatively
parallels the chemical freeze-out systematics of
Cleymans~et~al.~\cite{PhysRevC.73.034905} while lying systematically below
them in temperature, as expected for a later kinetic decoupling}.

A second important feature visible in Fig.~\ref{fig:t-mub} is the systematic 
upward shift in the extracted $T$ as $v_z$ increases, at all collision 
energies. This effect is not a consequence of longitudinal flow hardening the 
$p_T$ spectra; longitudinal motion is along the beam axis and cannot
directly transfer momentum into the transverse plane. Rather, the origin is 
kinematic: in our model, the distribution function depends on the 
Lorentz-invariant product $p^\mu u_\mu$, where $u^\mu$ is the four-velocity 
of the fluid element. When $v_z \neq 0$, the longitudinal component of 
$u^\mu$ is nonzero, and the effective energy of a particle as seen in the 
fluid rest frame receives a contribution from the longitudinal motion through 
the Lorentz factor $\gamma_z = (1 - v_z^2)^{-1/2}$. Consequently, at fixed 
observed $p_T$ and rapidity $y$, the argument of the thermal distribution is 
modified relative to the $v_z = 0$ case. To reproduce the same observed 
spectrum, the least-squares fit compensates by returning a higher value of
$T$. \fx{Because the rapidity window is narrow,
$\cosh(y-y_{\parallel}^{\prime}) \simeq \cosh y_{\parallel}^{\prime} =
\gamma_z$ across it, so the exponent of Eq.~\eqref{Eq:eq15} depends on $v_z$
only through the ratio $T/\gamma_z$ and the fit should return
$T(v_z)\simeq\gamma_z T(0)$. Table~\ref{table:fitparams} bears this out at
every energy: $T(v_z)/T(0)$ exceeds $\gamma_z$ by only $0.3\%$ at $v_z=0.2$
and $1.5$--$2.0\%$ at $v_z=0.4$, the residual coming from the accompanying
rise in $v_T$.} This represents a genuine degeneracy in the model between the
longitudinal collective velocity and the thermal temperature, and shows
the importance of \fx{stating $v_z$ explicitly: a temperature quoted without
it is an effective temperature that has silently absorbed the longitudinal
flow. Since the spectra themselves cannot discriminate between the three $v_z$
(the $\chi^2$ degrades by only $2$--$6\%$ from $v_z=0$ to $0.4$, as noted
above), $v_z$ is held fixed and scanned here rather than fitted.} \rev{The
baryon chemical potential is $v_z$-independent by construction here, since it
is fixed to the Cleymans systematics of Eq.~\eqref{eq:mubS}; this is
consistent with the physical expectation that the net-baryon content at
mid-rapidity is governed primarily by the initial stopping dynamics rather than
by the subsequent collective expansion. Crucially, over the explored range
$v_z \leq 0.4$ the shift leaves the fitted $T$ at or below the crossover band
$T_c \approx 155$--$160$ MeV at every energy; the extrapolation of this trend
to larger $v_z$ is quantified in Sec.~\ref{Section:corr}.}

\fx{Before presenting the densities, a comment is in order on the meaning of
the $(\rho_B,\varepsilon)$ points extracted here.
Ref.~\cite{PhysRevC.74.047901} evaluated the HRG densities at the
\emph{chemical} freeze-out point $(T_{\rm ch},\mu_B^{\rm ch})$. Our spectral
fits instead determine the \emph{kinetic} freeze-out temperature, while the
baryon chemical potential is supplied externally by the chemical freeze-out
systematics, Eq.~\eqref{eq:mubS}. Evaluating an equilibrium HRG at the hybrid
point $(T_{\rm kin},\mu_B^{\rm ch})$ is therefore an approximation: between
chemical and kinetic freeze-out the abundances are frozen and the hadron gas
is only in \emph{partial} chemical equilibrium, with effective chemical
potentials that grow as the system cools~\cite{Bebie1992,Hirano2002}. Since
$\mu_B/T$ increases during this (approximately isentropic) hadronic cooling,
the densities reported below should be read as conservative (lower) estimates
of the net-baryon compression at kinetic decoupling; this also explains
naturally why our $\rho_B$ values lie below the chemical-surface maximum of
Ref.~\cite{PhysRevC.74.047901}. A self-consistent treatment, enforcing
conservation of the frozen yields (e.g.\ constant $\rho_B/s$ from the chemical
surface down to $T_{\rm kin}$), is left for future work; a first step in this
direction, the data-driven determination of the net-baryon chemical potential
on the kinetic surface itself from the $\bar p/p$ ratio, is presented in
Sec.~\ref{Section:mubextraction}.}

Using the extracted $T$ and $\mu_B$ at each collision energy, we compute
the net baryon density via Eq.~\eqref{Eq:netBaryon} and the energy density 
via Eq.~\eqref{Eq:energyDen2}, setting $\mu_S = \mu_Q = 0$ and employing
\rev{the full Bose--Einstein/Fermi--Dirac series (summed to convergence). }
The results are displayed in Fig.~\ref{fig:t-rhoE} and tabulated in
Table~\ref{table:epsilon}.

\begin{table}[h!]
\centering
\begin{tabular}{||c c c c||}
 \hline
 $\sqrt{s_{NN}}$ & $v_z$ & $\rho_{B}$ & $\varepsilon$ \\
 (GeV) & & (fm$^{-3}$) & (MeV/fm$^{3}$) \\ [0.5ex]
 \hline\hline
      & 0.0 & \rev{0.0535} & \rev{120.1} \\
 7.7  & 0.2 & \rev{0.0642} & \rev{143.2} \\
      & 0.4 & \rev{0.1187} & \rev{261.9} \\
 \hline
      & 0.0 & \rev{0.0193} & \rev{68.5}  \\
 11.5 & 0.2 & \rev{0.0236} & \rev{81.3}  \\
      & 0.4 & \rev{0.0465} & \rev{147.8} \\
 \hline
      & 0.0 & \rev{0.0216} & \rev{120.3} \\
 19.6 & 0.2 & \rev{0.0267} & \rev{143.2} \\
      & 0.4 & \rev{0.0540} & \rev{262.6} \\
 \hline
      & 0.0 & \rev{0.0149} & \rev{115.8} \\
 27   & 0.2 & \rev{0.0186} & \rev{137.9} \\
      & 0.4 & \rev{0.0388} & \rev{254.2} \\
 \hline
      & 0.0 & \rev{0.0115} & \rev{125.4} \\
 39   & 0.2 & \rev{0.0144} & \rev{150.0} \\
      & 0.4 & \rev{0.0309} & \rev{281.6} \\ [1ex]
 \hline
\end{tabular}
\caption{\fx{Net baryon density $\rho_B$ and energy density $\varepsilon$ on
the kinetic freeze-out surface, for each collision energy and each fixed
longitudinal velocity $v_z$. They are obtained by evaluating the hadron
resonance gas, Eqs.~\eqref{Eq:numerDen2}, \eqref{Eq:netBaryon}
and~\eqref{Eq:energyDen2}, at the fitted temperature $T$ of
Table~\ref{table:fitparams} and the Cleymans input $\mu_B$ of
Eq.~\eqref{eq:mubS}, with $\mu_S = \mu_Q = 0$ and the full
Bose--Einstein/Fermi--Dirac series summed to convergence.
$\mu_B$ depends only on $\sqrt{s_{NN}}$, so the
$v_z$ dependence enters entirely through $T$. These are the points plotted in
Fig.~\ref{fig:t-rhoE}. This is the hybrid $(T_{\rm kin},\mu_B^{\rm ch})$
evaluation discussed above; the fully data-driven densities are given in
Sec.~\ref{Section:mubextraction}.}}
\label{table:epsilon}
\end{table}

The net baryon density is largest at $\sqrt{s_{NN}} = 7.7$ GeV and decreases
\rev{overall} with increasing collision energy. This behavior directly
reflects the energy dependence of $\mu_B$: as $\sqrt{s_{NN}}$ increases,
the decreasing degree of baryon stopping leads to a smaller net-baryon
density at mid-rapidity, and correspondingly smaller $\mu_B$. \rev{The mild
non-monotonicity around $\sqrt{s_{NN}} = 11.5$--$19.6$ GeV mirrors the
correspondingly lower fitted $T$ at $11.5$ GeV, to which $\rho_B$ is sensitive
through the Boltzmann factor $e^{-m/T}$.} \fx{The underlying temperature
difference ($\sim 2$~MeV) is well within the fit uncertainties, so this dip is
not statistically significant.} At higher
collision energies, freeze-out therefore occurs in a more baryon-dilute
environment characterized by low $\mu_B$ and low net baryon density, but
at a higher thermal temperature\fx{, paralleling the qualitative trend of the
chemical freeze-out systematics at correspondingly lower (kinetic)
temperatures}.

\rev{The extracted freeze-out temperatures span the ranges \fx{$119$}--$136$ MeV for
$v_z = 0.0$, $122$--$140$ MeV for $v_z = 0.2$, and $132$--$152$ MeV for
$v_z = 0.4$, across the collision energy range $\sqrt{s_{NN}} = 7.7$--$39$
GeV. The systematic upward shift of approximately $30$--$40$ MeV per unit
increment in $v_z$ is consistent with the kinematic degeneracy between
longitudinal collective motion and thermal temperature discussed above: the
larger the longitudinal velocity of the fireball, the larger the compensation
in $T$ required by the fit to reproduce the observed spectra. Importantly,
for the explored range $v_z \leq 0.4$ these temperatures all lie at or below
the lattice-QCD crossover band $T_c \approx 155$--$160$ MeV, so the
hadron-resonance-gas description underlying the density evaluation remains
internally consistent at every energy and every $v_z$ considered here.}

\begin{figure*}[htb]
\centering
\includegraphics[width=0.95\textwidth]{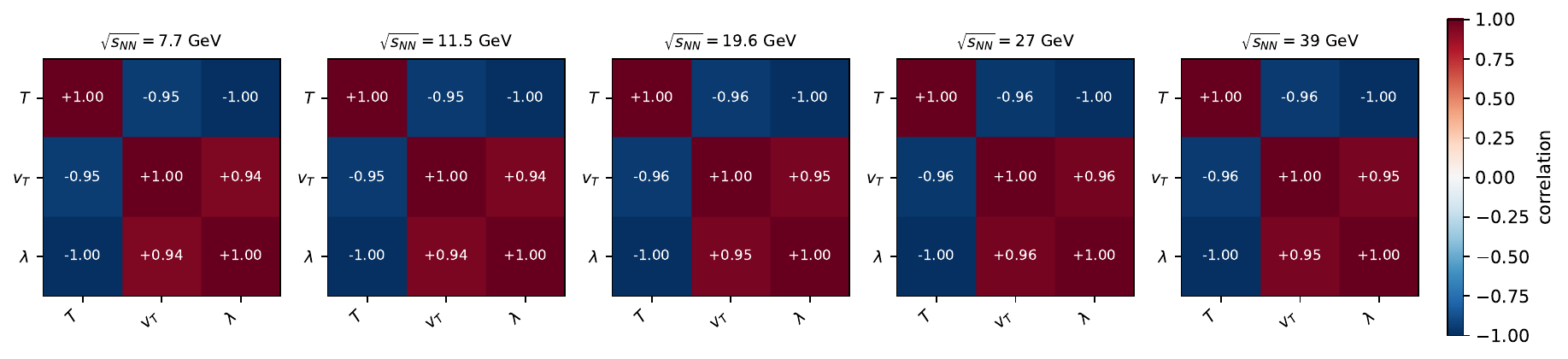}
\caption{\rev{(Color online) Fitted-parameter correlation matrices for the
proton fit $(T,v_T,\lambda)$ at $v_z=0$, for each STAR energy. The strong
$T$--$v_T$ anti-correlation ($\approx-0.95$) and $T$--$\lambda$
anti-correlation ($\approx-1.00$) are essentially energy-independent.}}
\label{fig:corrmatrix}
\end{figure*}

\begin{figure}[htb]
\centering
\includegraphics[width=0.48\textwidth]{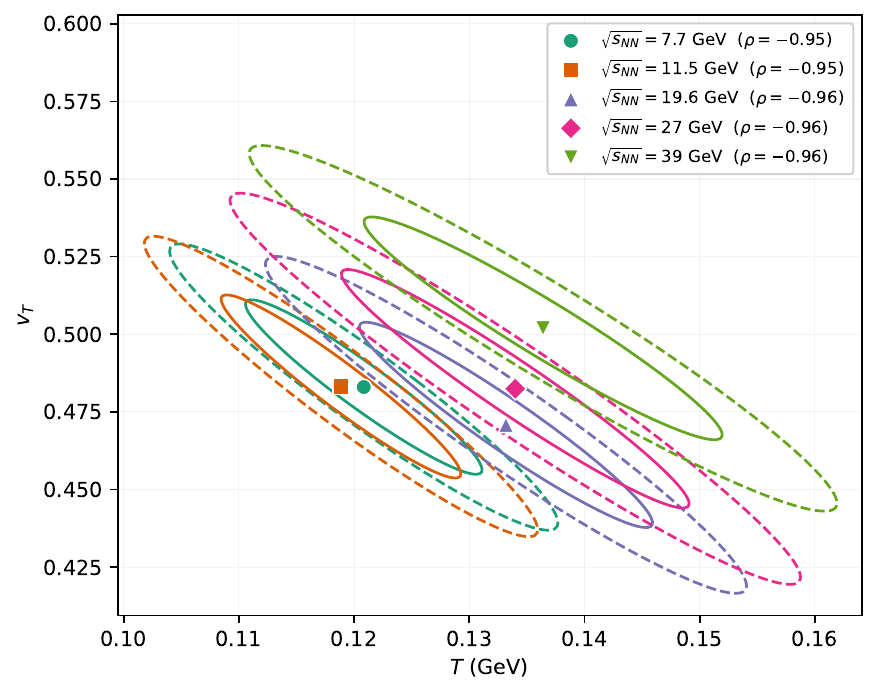}
\caption{\rev{(Color online) Joint $1\sigma$ (solid) and $2\sigma$ (dashed)
confidence regions in the $(T,v_T)$ plane for $v_z=0$. The narrow, tilted
ellipses are the geometric expression of the $T$--$v_T$ anti-correlation:
$T$ and $v_T$ are jointly, not independently, constrained.} \fx{The contours
are curves of constant $\Delta\chi^2$ from the $(T,v_T)$ block of the fit
covariance, at the two-parameter levels $\Delta\chi^2 = 2.30$ and $6.18$
($68.27\%$ and $95.45\%$); as joint regions they project onto the $T$ axis as
$\sqrt{2.30} = 1.52$ times the one-parameter errors of
Table~\ref{table:fitparams}.}}
\label{fig:ellipse}
\end{figure}

\begin{figure}[htb]
\centering
\includegraphics[width=0.48\textwidth]{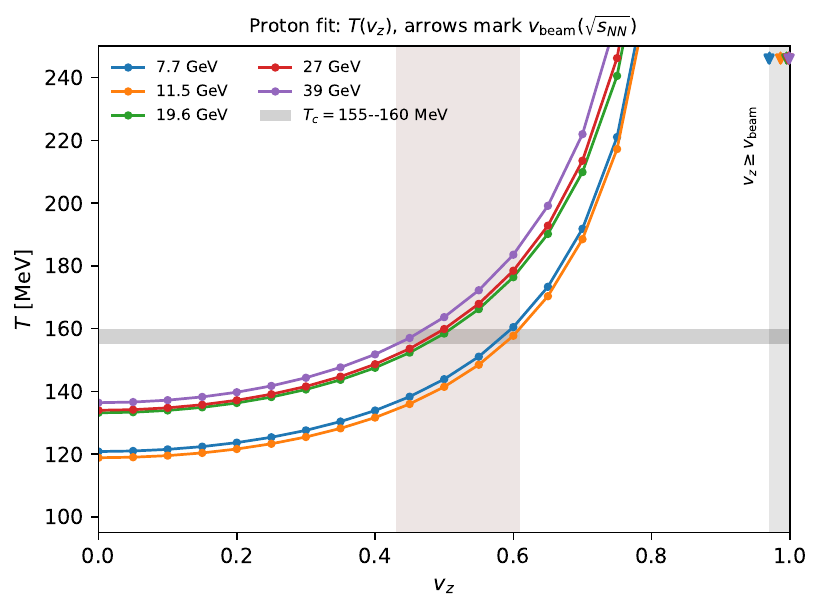}
\caption{\rev{(Color online) Fitted kinetic freeze-out temperature $T$ as a
function of the longitudinal flow velocity $v_z$, for the five STAR energies.
The grey band is the crossover temperature $T_c=155$--$160$~MeV; the vertical
shaded strip marks the range of critical values $v_z^{\,c}\approx0.43$--$0.61$
(Table~\ref{table:vzc}) at which $T$ crosses $T_c$.} \fx{The color-matched
downward arrows near $v_z = 1$ mark the kinematic ceiling
$v_{\rm beam}(\sqrt{s_{NN}})$ for each energy, and the shaded region beyond
$v_{\rm beam}(7.7~\mathrm{GeV}) = 0.970$ is kinematically inaccessible; the
thermodynamic bound $v_z^{\,c}$ is far more restrictive.}}
\label{fig:tvz}
\end{figure}

{
\subsection{Parameter correlations and a bound on the longitudinal flow}
\label{Section:corr}

Because the transverse-momentum slope can be reproduced either by a higher
temperature with weaker flow or by a lower temperature with stronger flow, the
fitted parameters are strongly correlated. We quantify this directly from the
fit covariance matrix. Table~\ref{table:corr} lists the Pearson correlation
coefficients between the free parameters $(T,v_T,\lambda)$ at each energy for
$v_z=0$, and Fig.~\ref{fig:corrmatrix} displays the full correlation matrices;
the pattern is essentially energy- and $v_z$-independent. Temperature and
transverse flow are strongly \emph{anti}-correlated,
$\rho_{T v_T}\approx-0.95$, the well-known thermal/flow ($T$--$\beta_T$)
degeneracy of blast-wave--type descriptions, while the normalization is almost
perfectly anti-correlated with $T$ ($\rho_{T\lambda}\approx-1.00$), which is the
same channel that makes $\mu_B$ undetermined once its fugacity is freed
(Sec.~\ref{Section:Results}).

\begin{table}[h!]
\centering
\renewcommand{\arraystretch}{1.15}
\begin{tabular}{||c c c c||}
 \hline
 $\sqrt{s_{NN}}$ (GeV) & $\rho_{T v_T}$ & $\rho_{T\lambda}$ & $\rho_{v_T\lambda}$ \\ [0.5ex]
 \hline\hline
 7.7  & $-0.950$ & $-0.998$ & $+0.937$ \\
 11.5 & $-0.954$ & $-0.998$ & $+0.943$ \\
 19.6 & $-0.955$ & $-0.999$ & $+0.948$ \\
 27   & $-0.963$ & $-0.999$ & $+0.957$ \\
 39   & $-0.957$ & $-0.999$ & $+0.951$ \\ [1ex]
 \hline
\end{tabular}
\caption{Fitted-parameter correlation coefficients for the proton fit at
$v_z=0$. The strong $T$--$v_T$ anti-correlation reflects the thermal/flow
degeneracy; $T$ and $v_T$ must be quoted jointly.}
\label{table:corr}
\end{table}

The practical consequence is that the one-parameter uncertainties on $T$ and
$v_T$ in Table~\ref{table:fitparams} understate the true joint uncertainty:
the allowed region in the $(T,v_T)$ plane is a narrow, steeply tilted ellipse,
shown in Fig.~\ref{fig:ellipse}. $T$ and $v_T$ should therefore be regarded as
jointly determined rather than independently measured, and claims about the
absolute value of either should be made with the correlation in mind. We note
in addition that the mid-rapidity spectra ($|y|<0.1$) carry very little
information on $v_z$ itself, since a longitudinal boost acts as a near-constant
rescaling within this narrow window; $v_z$ is therefore scanned as a fixed
external parameter rather than fitted.

Although $v_z$ cannot be determined from the mid-rapidity spectra, its effect
on the extracted temperature provides an indirect, thermodynamically motivated
upper bound. Fixing $\mu_B$ and refitting $(T,v_T,\lambda)$ over a fine grid in
$v_z$, we find that $T$ rises smoothly and monotonically with $v_z$
(Fig.~\ref{fig:tvz}), crossing the crossover band
$T_c\approx155$--$160$~MeV at a critical value $v_z^{\,c}$ listed in
Table~\ref{table:vzc}. Beyond $v_z^{\,c}$ the fitted kinetic temperature would
exceed $T_c$, i.e.\ the hadron-resonance-gas description on which the model
rests would no longer be self-consistent. The bound is
$v_z^{\,c}\approx0.57$--$0.61$ at the lowest energies and tightens to
$v_z^{\,c}\approx0.43$--$0.51$ at $39$~GeV, where the baseline temperature is
already higher. The values $v_z\le0.4$ explored above thus lie safely within
the physically admissible region at all energies, while longitudinal flow
velocities in excess of roughly half the speed of light are disfavored on the
grounds that they would push kinetic freeze-out above the crossover
temperature.

\fx{Two kinematic remarks complete this discussion. First, no point of the
scan is superluminal by construction: with the rapidity parametrization of
Eq.~\eqref{Eq:eq11} the total laboratory speed of a fluid element obeys
$|v|^2 = v_z^2 + (1 - v_z^2)\,v_T^{\prime\,2} < 1$ for any $v_z, v_T' < 1$
(relativistic velocity addition is built into the four-velocity), the largest
value reached anywhere on the grid being $|v| \simeq 0.98$. Second, the
longitudinal collective velocity of matter produced at mid-rapidity cannot
exceed the beam velocity in the center-of-mass frame,
$v_{\rm beam} = \tanh y_{\rm beam}$ with
$\gamma_{\rm beam} = \sqrt{s_{NN}}/(2m_N)$, which ranges from $0.970$ at
$7.7$~GeV to $0.999$ at $39$~GeV (Table~\ref{table:vzc}, arrows in
Fig.~\ref{fig:tvz}). Since $v_z^{\,c} \approx 0.43$--$0.61$ lies far below
this kinematic ceiling at every energy, the $T \le T_c$ requirement, not
kinematics, is the operative constraint on the longitudinal flow.}

\begin{table}[h!]
\centering
\renewcommand{\arraystretch}{1.15}
\begin{tabular}{||c c c c||}
 \hline
 $\sqrt{s_{NN}}$ (GeV) & $v_z^{\,c}$ ($T_c{=}155$) & $v_z^{\,c}$ ($T_c{=}160$) & \fx{$v_{\rm beam}$} \\ [0.5ex]
 \hline\hline
 7.7  & 0.57 & 0.60 & \fx{0.970} \\
 11.5 & 0.59 & 0.61 & \fx{0.987} \\
 19.6 & 0.47 & 0.51 & \fx{0.995} \\
 27   & 0.46 & 0.50 & \fx{0.998} \\
 39   & 0.43 & 0.47 & \fx{0.999} \\ [1ex]
 \hline
\end{tabular}
\caption{Critical longitudinal velocity $v_z^{\,c}$ at which the fitted kinetic
freeze-out temperature reaches the lower ($155$~MeV) and upper ($160$~MeV)
edges of the crossover band, obtained by interpolating the $T(v_z)$ scan of
Fig.~\ref{fig:tvz}. \fx{The last column lists the kinematic ceiling: the beam
velocity in the center-of-mass frame, $v_{\rm beam} = \tanh y_{\rm beam}$ with
$\gamma_{\rm beam} = \sqrt{s_{NN}}/(2m_N)$, which the fireball's longitudinal
collective velocity cannot exceed. The thermodynamic bound $v_z^{\,c}$ is far
more restrictive than the kinematic one at all energies.}}
\label{table:vzc}
\end{table}
}

\begin{figure}[htb]
\centering
\includegraphics[width=0.47\textwidth]{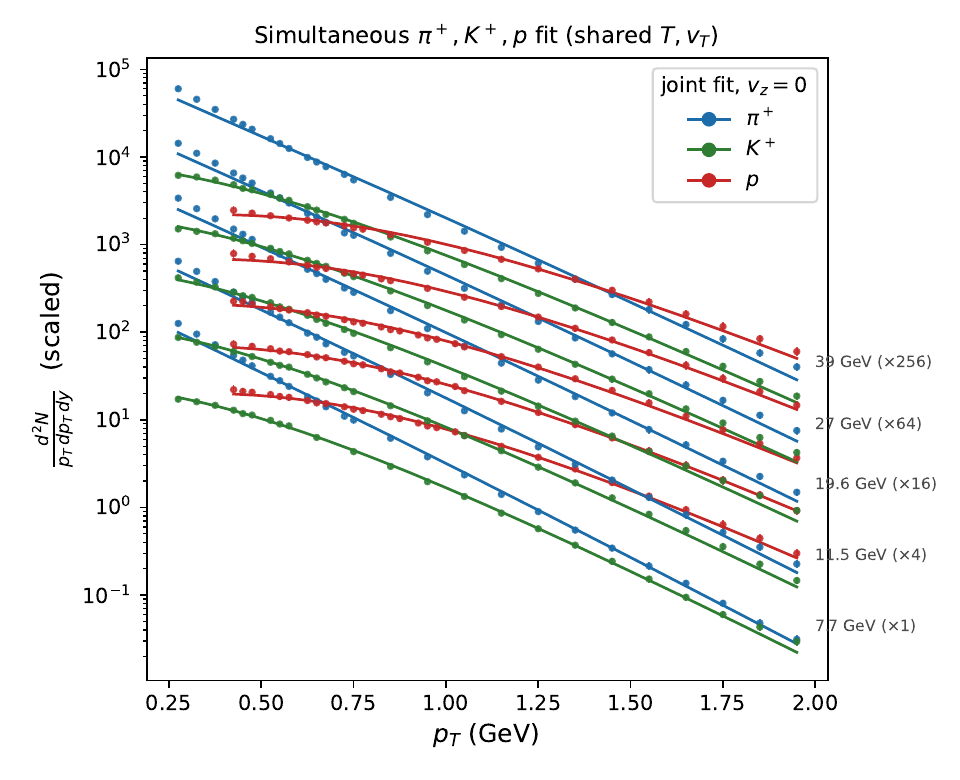}
\caption{\nw{(Color online) Simultaneous fit of the $\pi^+$ (blue), $K^+$
(green), and proton (red) $p_T$ spectra with a shared kinetic freeze-out
temperature $T$ and transverse flow $v_T$ (curves), at $v_z=0$ for the five STAR
energies (successively scaled by powers of 4 for clarity). $\mu_B$ is fixed to
the Cleymans systematics.}}
\label{fig:msspectra}
\end{figure}

\begin{figure*}[htb]
\centering
\includegraphics[width=0.92\textwidth]{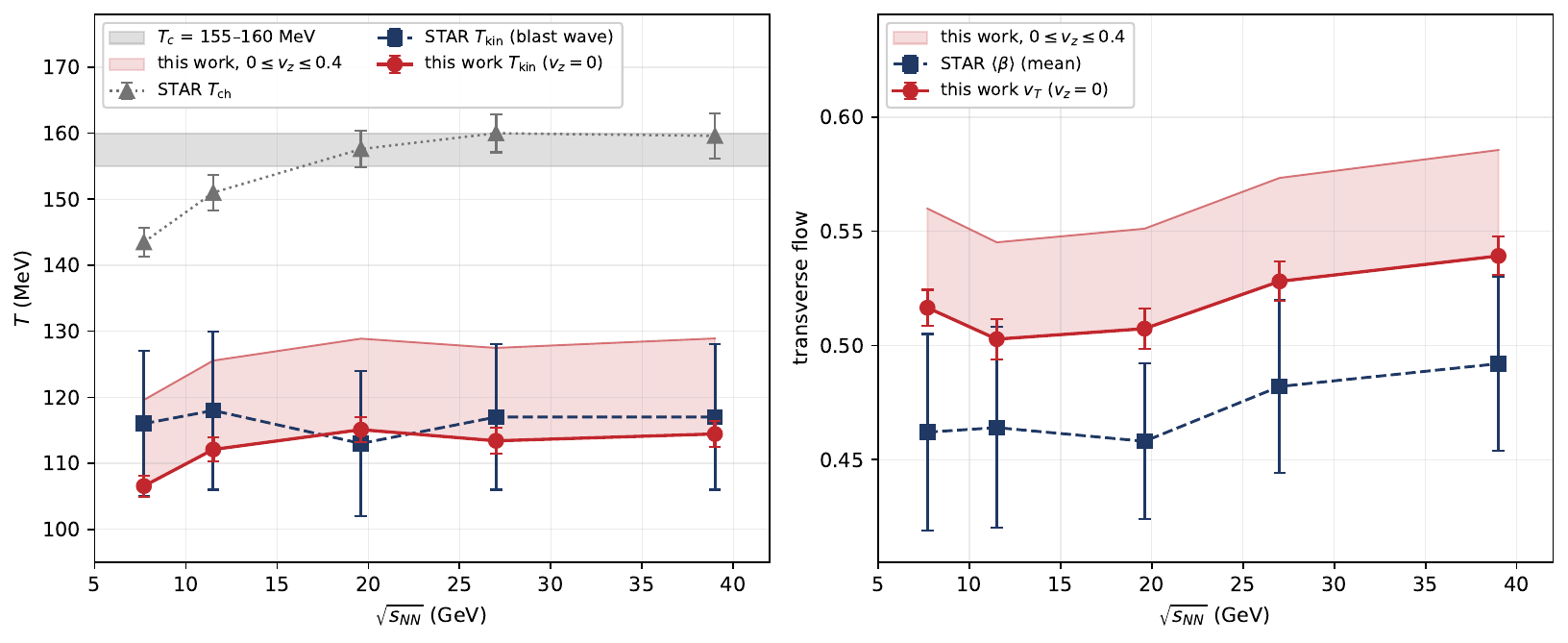}
\caption{\nw{(Color online) Left: the multi-species kinetic freeze-out
temperature $T_{\rm kin}$ (this work, $v_z=0$) compared with STAR's published
blast-wave $T_{\rm kin}$ and chemical-freeze-out $T_{\rm ch}$; the grey band is
$T_c=155$--$160$~MeV. Right: our \fx{single} flow \fx{velocity} $v_T$ compared
with STAR's mean
transverse velocity $\langle\beta\rangle$. STAR values from
Ref.~\cite{PhysRevC.96.044904}.} \fx{All error bars are $1\sigma$: ours from the
fit covariance, STAR's their published combined statistical and systematic
uncertainties (Tables~IX and X of Ref.~\cite{PhysRevC.96.044904}). The shaded
red bands span the scanned range $0\le v_z\le0.4$ and indicate the systematic
associated with the longitudinal flow, which the mid-rapidity spectra do not
constrain; the plotted points are $v_z=0$.}}
\label{fig:starcompare}
\end{figure*}

{
\subsection{Simultaneous multi-species fit and cross-check against STAR}
\label{Section:multispecies}

The strong $T$--$v_T$ anti-correlation of the single-species fit
(Sec.~\ref{Section:corr}) originates in the fact that the same spectral slope
can be produced by a hotter, less rapidly flowing source or by a cooler, faster
one. This ambiguity is lifted by the particle mass: under a common radial flow,
heavier hadrons receive a larger transverse blue-shift, so the spectra of
species of different mass constrain $T$ and $v_T$ in different combinations.
We therefore perform a \emph{simultaneous} fit of the $\pi^+$, $K^+$, and
proton $p_T$ spectra at each energy, sharing a single kinetic freeze-out
temperature $T$ and transverse flow velocity $v_T$ while allowing an
independent normalization for each species, with $\mu_B$ again fixed to the
Cleymans systematics. Because the fitted observables are the spectral
\emph{shapes} and the flow, this remains a determination of the \emph{kinetic}
freeze-out surface; the chemical information ($\mu_B$) is supplied externally.
\fx{As shown above, rescaling that external $\mu_B$ by factors of four in
either direction leaves the fitted $T$ and $v_T$ of this three-species fit
unchanged to better than $3\times10^{-5}$~MeV and $10^{-7}$ respectively, so
the chemical input is inert here as well.}

\fx{It is the sharing of a single $T$ and $v_T$ across species of different
mass, not the freedom of the individual normalizations, that removes the
near-singular $T$--$\lambda$ direction of the single-species fit: a change in
$T$ can still be absorbed by the normalization of any one species, but the
compensation required depends on the particle mass, so it cannot be made for
$\pi^+$, $K^+$ and $p$ at once.} The joint constraint tightens accordingly:
the resulting $1\sigma$ uncertainties on
$T$ and $v_T$ shrink substantially relative to the proton-only fit,
\fx{$\sigma_T$ falling from $6.8$--$10.3$~MeV to $1.6$--$2.0$~MeV and
$\sigma_{v_T}$ from $0.019$--$0.025$ to $0.008$--$0.009$, gains of factors
$3.7$--$5.3$ and $2.2$--$2.9$ respectively
(Tables~\ref{table:fitparams} and~\ref{table:starcompare})}.
The shared kinetic temperature is correspondingly lower and more physical,
$T_{\rm kin}\simeq107$--$115$~MeV at $v_z=0$ (rising to $120$--$129$~MeV at
$v_z=0.4$), and is nearly independent of $\sqrt{s_{NN}}$. The residual
$T$--$v_T$ anti-correlation ($\rho\approx-0.94$) persists, as it must for any
blast-wave--type description, but the joint confidence region is far smaller.

Crucially, these values agree with STAR's own published kinetic freeze-out
parameters, obtained from an independent blast-wave analysis of the same
data~\cite{PhysRevC.96.044904}. As shown in Table~\ref{table:starcompare} and
Fig.~\ref{fig:starcompare}, our $T_{\rm kin}$ agrees with the STAR values at
every energy: \fx{the differences run from $+2.1$ to $-9.5$~MeV and remain
within $0.9\sigma$ of the combined uncertainties throughout, STAR's blast-wave
temperatures themselves carrying $\pm11$--$12$~MeV.} Our $T_{\rm kin}$ lies
well below the STAR chemical
freeze-out temperature $T_{\rm ch}\simeq144$--$160$~MeV, exactly as expected if
kinetic freeze-out occurs later and cooler than chemical freeze-out. Our fitted
$v_T$ lies \fx{systematically above} STAR's mean transverse velocity
$\langle\beta\rangle$ \fx{by $0.04$--$0.06$ ($0.9$--$1.4\sigma$)}, \fx{as
expected for} a single
flow velocity \fx{compared with the average over} a velocity profile. Finally,
the external Cleymans
$\mu_B$ we adopt \fx{lies $5$--$8\%$ above} STAR's measured chemical-freeze-out
$\mu_B$ \fx{at every energy}. The simultaneous $\pi^+/K^+/p$ fit and
its fit curves are shown in Fig.~\ref{fig:msspectra}.

\fx{Figure~\ref{fig:starcompare} displays the comparison. In the left panel our
shared kinetic temperature tracks STAR's blast-wave $T_{\rm kin}$ across the
whole BES range, while both lie far below the chemical freeze-out temperature,
which rises from $143.5$~MeV at $7.7$~GeV and levels off inside the crossover
band from $19.6$~GeV upward ($157.6$, $160.0$ and $159.6$~MeV). The separation
$T_{\rm ch}-T_{\rm kin}$ consequently widens with energy, from $37$~MeV at
$7.7$~GeV to $45$--$47$~MeV at $27$--$39$~GeV, as expected if the hadronic phase
between the two freeze-out surfaces is longer-lived at higher collision energy.
The right panel shows our $v_T$ and STAR's $\langle\beta\rangle$ rising together
with $\sqrt{s_{NN}}$, the offset between them staying nearly constant at
$0.04$--$0.06$. The shaded bands span the scanned range $0\le v_z\le0.4$: within
STAR's uncertainties the comparison is insensitive to $v_z$, so it corroborates
the kinetic interpretation of the fit without independently constraining the
longitudinal flow.}

\begin{table}[h!]
\centering
\renewcommand{\arraystretch}{1.15}
\begin{tabular}{||c cc c cc||}
 \hline
 $\sqrt{s_{NN}}$ & $T_{\rm kin}$ & STAR & $v_T$ & STAR & STAR \\
 (GeV) & (this work) & $T_{\rm kin}$ & (this work) & $\langle\beta\rangle$ & $\mu_B$ \\
  & (MeV) & (MeV) & & & (MeV) \\
 \hline\hline
 7.7  & \fx{106.5(16)} & \fx{116(11)} & \fx{0.517(8)} & \fx{0.462(43)} & 400 \\
 11.5 & \fx{112.1(19)} & \fx{118(12)} & \fx{0.503(9)} & \fx{0.464(44)} & 292 \\
 19.6 & \fx{115.1(19)} & \fx{113(11)} & \fx{0.507(9)} & \fx{0.458(34)} & 194 \\
 27   & \fx{113.4(19)} & \fx{117(11)} & \fx{0.528(9)} & \fx{0.482(38)} & 149 \\
 39   & \fx{114.4(20)} & \fx{117(11)} & \fx{0.539(8)} & \fx{0.492(38)} & 104 \\ [1ex]
 \hline
\end{tabular}
\caption{Kinetic freeze-out temperature $T_{\rm kin}$ and transverse flow $v_T$
from the simultaneous $\pi^+/K^+/p$ fit at $v_z=0$, compared with STAR's
published blast-wave $T_{\rm kin}$ and $\langle\beta\rangle$ and measured
chemical-freeze-out $\mu_B$~\cite{PhysRevC.96.044904}. The Cleymans $\mu_B$ used
as input (Table~\ref{table:fitparams}) \fx{lies $5$--$8\%$ above the STAR
$\mu_B$ at every energy}. \fx{Numbers in parentheses are $1\sigma$
uncertainties in units of the last digit. Ours are the one-parameter errors
from the fit covariance, given unrounded so that they may be compared directly
with the proton-only values of Table~\ref{table:fitparams}; the STAR columns
carry that collaboration's published combined statistical and systematic
errors.}}
\label{table:starcompare}
\end{table}

Applying the $T\le T_c$ criterion of Sec.~\ref{Section:corr} to the more robust
multi-species temperature refines the bound on the longitudinal flow. Scanning
the joint fit in $v_z$ (Fig.~\ref{fig:mstvz}), the shared $T$ reaches the
crossover band at a critical value $v_z^{\,c}\approx0.58$--$0.65$, somewhat
larger than the proton-only estimate because the multi-species $T_{\rm kin}$ is
lower. The explored range $v_z\le0.4$ therefore lies comfortably within the
physically admissible region at all energies, and longitudinal flow velocities
above $v_z\approx0.6$ are disfavored on the grounds that they would drive
kinetic freeze-out above the hadron-resonance-gas limit.
}

\begin{figure}[htb]
\centering
\includegraphics[width=0.47\textwidth]{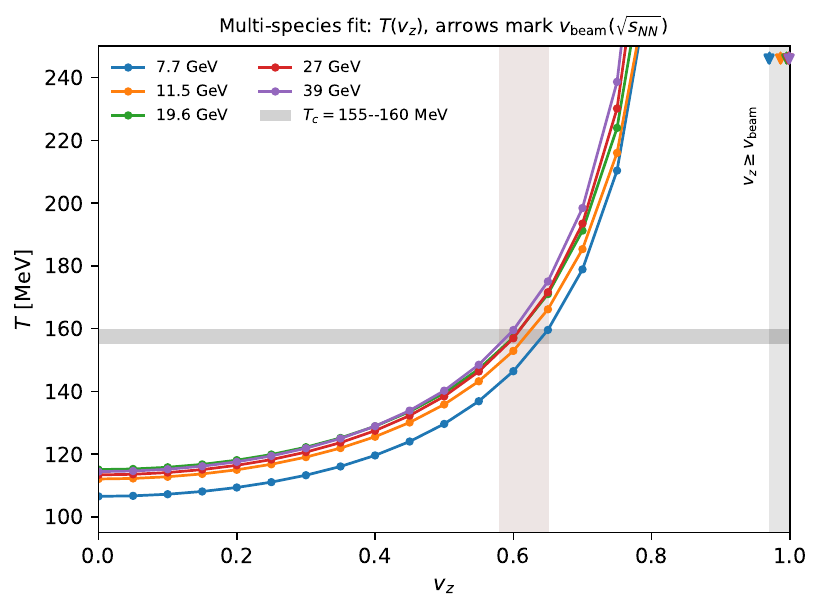}
\caption{\nw{(Color online) Shared kinetic freeze-out temperature from the
simultaneous $\pi^+/K^+/p$ fit as a function of $v_z$. The grey band is
$T_c=155$--$160$~MeV; the vertical shaded strip marks the refined critical range
$v_z^{\,c}\approx0.58$--$0.65$.} \fx{Arrows near $v_z = 1$ mark the kinematic
ceiling $v_{\rm beam}$ per energy, as in Fig.~\ref{fig:tvz}.}}
\label{fig:mstvz}
\end{figure}

\begin{figure*}[htb]
\centering
\includegraphics[width=0.95\textwidth]{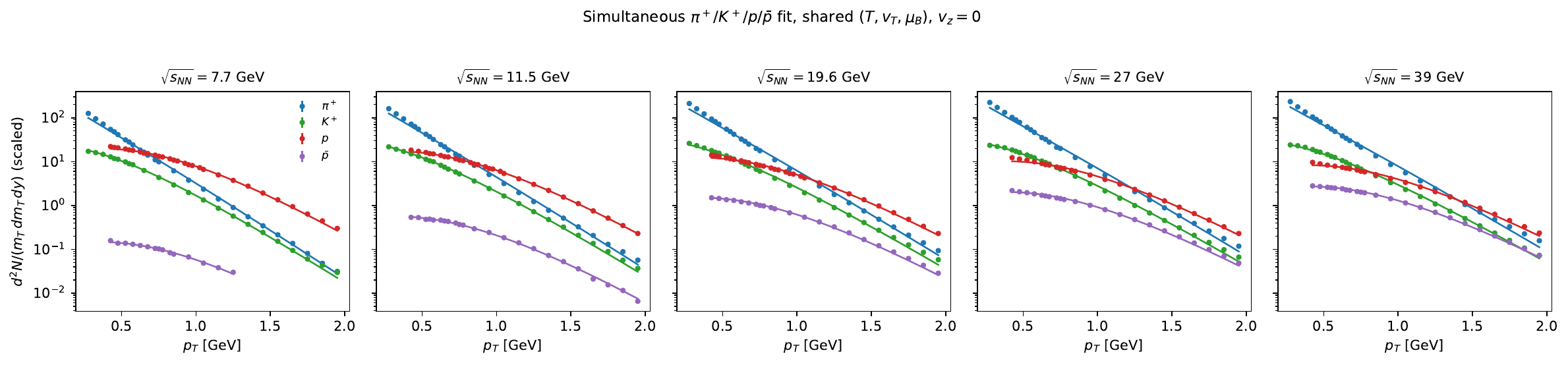}
\caption{\fx{(Color online) Simultaneous fit of the $\pi^+$ (blue), $K^+$
(green), $p$ (red), and $\bar p$ (purple) $p_T$ spectra with shared
$(T, v_T, \mu_B)$ and a common $p$/$\bar p$ normalization, at $v_z = 0$ for
the five STAR energies. The $\bar p/p$ splitting is controlled entirely by
the fugacity factor $e^{\pm\mu_B/T}$, which fixes the ratio $\mu_B/T$; the
simultaneously fitted $\pi^+$ and $K^+$ spectra, which are insensitive to
$\mu_B$ but share the same $T$, determine the temperature, and the two together
yield $\mu_B$.}}
\label{fig:v4spectra}
\end{figure*}

\begin{figure*}[htb]
\centering
\includegraphics[width=0.85\textwidth]{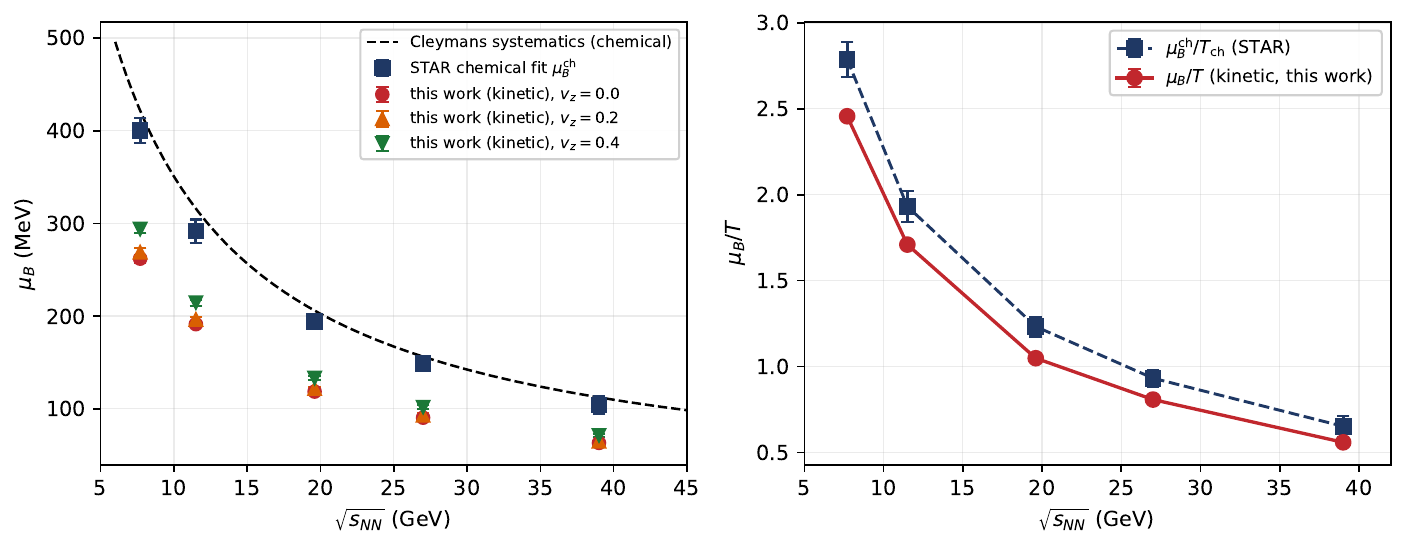}
\caption{\fx{(Color online) Left: the data-driven kinetic-surface $\mu_B$
(circles/triangles, three $v_z$ values) compared with the Cleymans chemical
freeze-out systematics (dashed) and STAR's measured chemical $\mu_B$
(squares)~\cite{PhysRevC.96.044904}. Right: the fitted $\mu_B/T$ ($v_z = 0$)
compared with STAR's chemical $\mu_B^{\rm ch}/T_{\rm ch}$; the near-agreement
reflects the conservation of the $\bar p/p$ ratio between the chemical and
kinetic surfaces. All error bars are $1\sigma$: ours from the fit covariance,
including the $T$--$\mu_B$ correlation, which cancels strongly in the ratio;
STAR's their published combined statistical and systematic uncertainties
(Table~IX of Ref.~\cite{PhysRevC.96.044904}), propagated to
$\mu_B^{\rm ch}/T_{\rm ch}$ treating $T_{\rm ch}$ and $\mu_B^{\rm ch}$ as
uncorrelated. The reference uncertainties dominate the comparison, and their
growth relative to the offset at the highest energy is why the deficit falls
from $\simeq 3\sigma$ below $27$~GeV to $1.5\sigma$ at $39$~GeV.}}
\label{fig:v4mub}
\end{figure*}

{
\subsection{Data-driven extraction of $\mu_B$ from the $\bar p/p$ ratio}
\label{Section:mubextraction}

The analyses above supply $\mu_B$ externally, because for any set of species
fitted with \emph{independent} normalizations the fugacity $e^{\mu_B B/T}$ is
absorbed into the factor $\mathcal{N}$ of each baryonic species,
Eq.~\eqref{Eq:lambdadef}, and $\mu_B$ remains
undetermined. The antiproton resolves this. We extend the simultaneous fit of
Sec.~\ref{Section:multispecies} to the four species $\pi^+$, $K^+$, $p$, and
$\bar p$, sharing a single kinetic temperature $T$, transverse flow $v_T$, and
baryon chemical potential $\mu_B$, with independent normalizations for
$\pi^+$ and $K^+$ but a \emph{common} volume factor $\mathcal{N}_p$ for $p$ and
$\bar p$. Since the proton and antiproton then differ only through their
fugacities $e^{\pm\mu_B/T}$, the measured $\bar p/p$ ratio determines $\mu_B$
directly, with no external chemical input ($\mu_S = \mu_Q = 0$ as before; the
$\bar p$ spectra are fitted over the STAR window
$0.425 \leq p_T \leq 1.95$~GeV, restricted to $p_T \leq 1.25$~GeV at
$7.7$~GeV by the data coverage). The fit quality remains good,
$\chi^2/\mathrm{NDF} \simeq 0.6$--$1.3$ (Fig.~\ref{fig:v4spectra}), \fx{though
the $\chi^2$ is not evenly distributed among the species: the $\pi^+$ spectra
carry a $\chi^2$ per point of $1.5$--$3.0$ and account for $63$--$76\%$ of the
total, the $K^+$ spectra $0.5$--$1.1$, and the $p$ and $\bar p$ spectra only
$0.1$--$0.4$. The temperature is therefore determined predominantly by the
meson spectra, as anticipated in Sec.~\ref{Section:multispecies}, while the
baryon spectra---reproduced well within their systematics-dominated
uncertainties---act mainly through the $\bar p/p$ ratio.} The
shared $T$ and $v_T$ move by at most $2$~MeV and $0.012$ respectively with
respect to the three-species fit, comparable to their $1\sigma$
uncertainties, confirming that the added $\bar p$ spectra are consistent
with the same kinetic surface.

The extracted chemical potential decreases from
$\mu_B = 262.0 \pm 4.2$~MeV at $\sqrt{s_{NN}} = 7.7$~GeV to
$63.2 \pm 1.7$~MeV at $39$~GeV ($v_z = 0$; Table~\ref{table:mubfit}), lying
systematically \emph{below} the chemical freeze-out values of
Ref.~\cite{PhysRevC.96.044904} ($\mu_B^{\rm ch} \simeq 400$--$104$~MeV,
Table~IX, strangeness-canonical yield fit) by \fx{approximately} the ratio of the
temperatures. This is precisely the behavior expected if the $\bar p/p$ ratio
is frozen at chemical freeze-out: since
$\bar p/p = e^{-2\mu_B/T}$, conservation of the ratio during the hadronic
cooling from $T_{\rm ch}$ to $T_{\rm kin}$ requires $\mu_B/T$ to remain
constant, i.e.\ $\mu_B^{\rm kin} \simeq (T_{\rm kin}/T_{\rm ch})\,
\mu_B^{\rm ch}$. Figure~\ref{fig:v4mub} shows that the extracted
$\mu_B/T = 2.46$--$0.56$ indeed tracks STAR's chemical
$\mu_B^{\rm ch}/T_{\rm ch} = 2.79$--$0.65$ at every energy, at the
$85$--$88\%$ level. Propagating the uncertainties quoted by STAR on
$T_{\rm ch}$ and $\mu_B^{\rm ch}$ \fx{together with our own fit covariance},
this deficit is significant at
\fx{$2.4$--$3.2\sigma$} for $\sqrt{s_{NN}} \leq 27$~GeV but only
\fx{$1.5\sigma$} at
$39$~GeV, where the chemical $\mu_B^{\rm ch}$ is itself known to
$\simeq 9\%$; the trend is therefore established by the lower energies. \fx{These
significances treat STAR's $T_{\rm ch}$ and $\mu_B^{\rm ch}$ uncertainties as
uncorrelated, which is conservative: the positive correlation between them in a
thermal fit would partially cancel in the ratio and increase the significance.}
The
residual offset is consistent with the neglected weak
feed-down (the inclusive STAR spectra receive different
$\Lambda$/$\bar\Lambda$ contributions for $p$ and $\bar p$, biasing the ratio
slightly upward and hence $\mu_B$ slightly downward) and with
$\mu_S = 0$.

\fx{Two features of the data warrant comment here. First, each published STAR
spectrum is a concatenation of two partially overlapping measurements covering
different $p_T$ intervals; in the overlap region the same particles are
therefore represented twice, and the two sets of points are treated here as
independent, which mildly over-weights that region. Second, and more
importantly, the sensitivity of the fit to the $p_T$ range exceeds what the
covariance errors alone would suggest. Refitting with only the higher-$p_T$
branch of each spectrum raises $T$ by $2$--$5$~MeV and changes $\mu_B$ by at
most $4$~MeV, within the quoted uncertainties; a fit restricted to the
lower-$p_T$ branch alone, however, returns $T \simeq 92$--$97$~MeV with
correspondingly smaller $\mu_B$. The latter is not an independent
determination: that branch spans only $p_T \lesssim 0.8$~GeV, precisely where
resonance feed-down, which the present thermal source does not include, is most
important, and where the $\pi^+$ residuals show a coherent trend rather than
random scatter. It does establish, however, that the dominant uncertainty on
the \emph{absolute scale} of $T$, and hence on $\mu_B$, is the model
description of the soft part of the spectrum rather than the fit statistics.
The ratio $\mu_B/T$, fixed by $\bar p/p$ alone, is markedly more robust: across
these fit variants it changes by at most $4.4\%$, against $\simeq 20\%$ for $T$
and $\mu_B$ separately. This is the quantity the data determine most directly,
and the stability of the $\mu_B/T$ comparison in Fig.~\ref{fig:v4mub} should be
read accordingly.}

The extracted $\mu_B/T$ is independent of the assumed $v_z$
(Table~\ref{table:mubfit}), as it must be, since the longitudinal boost
affects both species identically. In the language of partial chemical
equilibrium, the fitted $\mu_B$ is the \emph{net-baryon} (antisymmetric)
chemical potential on the kinetic surface, while the symmetric
over-abundance factors common to particles and antiparticles are absorbed
into the species normalizations.

\begin{table}[h!]
\centering
\renewcommand{\arraystretch}{1.15}
\begin{tabular}{||c c c c c c||}
 \hline
 $\sqrt{s_{NN}}$ & $T$ & $v_T$ & $\mu_B$ & $\mu_B/T$ & $\mu_B^{\rm ch}/T_{\rm ch}$ \\
 (GeV) & (MeV) & & (MeV) & & (STAR) \\ [0.5ex]
 \hline\hline
 7.7  & $106.6\pm1.5$ & $0.516\pm0.007$ & $262.0\pm4.2$ & 2.46 & 2.79 \\
 11.5 & $111.9\pm1.6$ & $0.503\pm0.007$ & $191.3\pm3.1$ & 1.71 & 1.93 \\
 19.6 & $113.1\pm1.6$ & $0.519\pm0.007$ & $118.6\pm2.2$ & 1.05 & 1.23 \\
 27   & $112.0\pm1.6$ & $0.537\pm0.006$ & $90.4\pm2.0$  & 0.81 & 0.93 \\
 39   & $113.1\pm1.6$ & $0.547\pm0.006$ & $63.2\pm1.7$  & 0.56 & 0.65 \\ [1ex]
 \hline
\end{tabular}
\caption{Simultaneous $\pi^+/K^+/p/\bar p$ fit at $v_z = 0$: shared kinetic
freeze-out temperature $T$, transverse flow $v_T$, and the \emph{data-driven}
baryon chemical potential $\mu_B$ extracted from the $\bar p/p$ ratio, with
$1\sigma$ uncertainties. The last two columns compare the fitted $\mu_B/T$
with STAR's chemical-freeze-out ratio $\mu_B^{\rm ch}/T_{\rm ch}$, taken from
the strangeness-canonical yield fit (SCEY) of Table~IX of
Ref.~\cite{PhysRevC.96.044904}; the grand-canonical yield fit (GCEY,
Table~VIII of the same reference) differs by $\leq 2\%$ (largest at $27$~GeV)
and would not alter any conclusion drawn here. At $v_z = 0.2$ ($0.4$) the temperatures rise by
$\simeq 3$ ($13$--$15$)~MeV and $\mu_B$ rises proportionally, leaving $\mu_B/T$
unchanged. $T$ and $\mu_B$ are strongly correlated
($\rho_{T\mu_B} = +0.93$ at $7.7$~GeV, falling to $+0.55$ at $39$~GeV), as
expected since the data constrain their ratio most tightly.}
\label{table:mubfit}
\end{table}

With both $T$ and $\mu_B$ now determined from the same spectra, the
$(\rho_B,\varepsilon)$ trajectory can be evaluated fully self-consistently on
the kinetic surface, with uncertainties propagated from the $(T,\mu_B)$
covariance (including their correlation). The result
(Fig.~\ref{fig:v4rhoeps}) is a substantially more dilute trajectory than the
hybrid $(T_{\rm kin},\mu_B^{\rm ch})$ evaluation of
Table~\ref{table:epsilon}: at $v_z = 0$ we find
$\rho_B = (4.4 \pm 0.8)\times10^{-3}$~fm$^{-3}$ and
$\varepsilon = 28 \pm 3$~MeV/fm$^3$ at $7.7$~GeV, falling to
$\rho_B = (0.90 \pm 0.15)\times10^{-3}$~fm$^{-3}$ at $39$~GeV, with
$\varepsilon$ approximately energy-independent at
$\simeq 28$--$34$~MeV/fm$^3$. The net baryon density now decreases
monotonically with $\sqrt{s_{NN}}$ (the non-monotonicity of the hybrid
evaluation, inherited from the statistically insignificant temperature dip at
$11.5$~GeV, is absent), while the dependence on the longitudinal flow persists
through the fitted temperature: \fx{at $v_z = 0.4$, $\rho_B$ is larger by a
factor of $\simeq 4$ and $\varepsilon$ by a factor of $\simeq 2.3$}. These
numbers characterize the dilute, cold endpoint of
the hadronic phase: within the equilibrium-HRG evaluation (which, as
discussed above, omits the symmetric partial-chemical-equilibrium
over-abundances and therefore provides lower estimates), kinetic decoupling
at the lowest BES energies occurs at a net baryon density of only a few
percent of normal nuclear density, a factor of $\simeq25$ below the
chemical-surface compression maximum of Ref.~\cite{PhysRevC.74.047901}.
}

\begin{figure}[htb]
\centering
\includegraphics[width=0.47\textwidth]{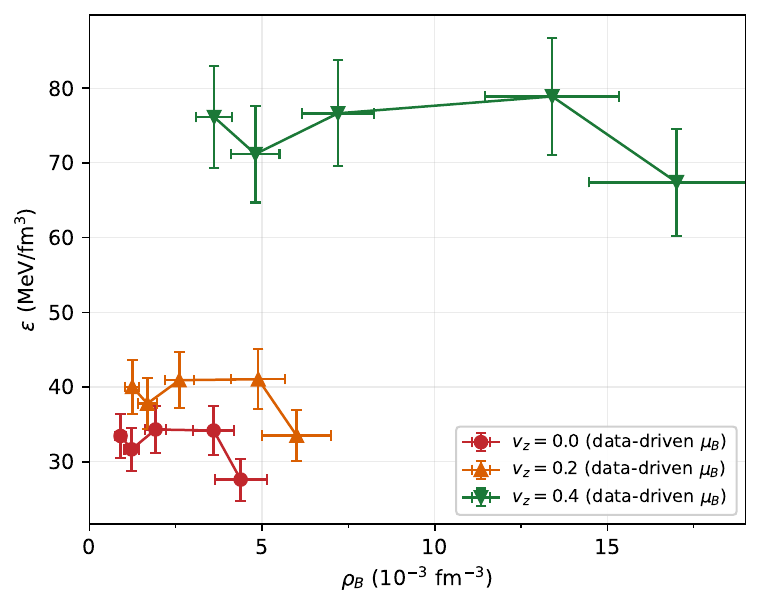}
\caption{\fx{(Color online) Fully data-driven kinetic freeze-out trajectory in
the $(\rho_B,\varepsilon)$ plane from the $\pi^+/K^+/p/\bar p$ fit, with
$1\sigma$ uncertainties propagated from the $(T,\mu_B)$ covariance, for the
three longitudinal velocities.}}
\label{fig:v4rhoeps}
\end{figure}

\vspace{0.5cm}

\section{Conclusions and Outlook}
\label{Section:conclusion}

\fx{In this study we have investigated kinetic freeze-out conditions in Au+Au
collisions at $\sqrt{s_{NN}} = 7.7$--$39$~GeV using a covariant statistical
fireball model~\cite{Touschek} that explicitly incorporates transverse and
longitudinal collective flow, extending to the kinetic freeze-out surface the
$(\rho_B,\varepsilon)$ mapping introduced by Randrup and
Cleymans~\cite{PhysRevC.74.047901}. Fitting STAR transverse-momentum spectra in
$0$--$5\%$ central collisions within $|y| < 0.1$, the analysis proceeds in three
stages of increasing constraint: a single-species proton fit and a three-species
$\pi^+/K^+/p$ fit, in both of which the baryon chemical potential $\mu_B$ must be
supplied externally, and finally a four-species fit including the antiproton, in
which $\mu_B$ becomes data-driven and no external chemical input enters the
analysis at all.}

Our principal findings are as follows. \rev{Because a single-species
mid-rapidity spectrum determines only the product of the normalization and the
fugacity, $\mu_B$ is not constrained by the fit; we therefore fix it to the
Cleymans systematics and fit only the temperature $T$, the transverse flow
$v_T$, and the normalization $\lambda$.} \fx{That choice is inert: rescaling the
fixed $\mu_B$ by a factor of four leaves the fitted $T$ and $v_T$ unchanged to
numerical precision, so the temperatures quoted here carry no
chemical-freeze-out assumption.} Longitudinal flow shifts the extracted
temperature systematically upward, \rev{the fitted $T$ spanning $119$--$136$ MeV
for $v_z = 0$, $122$--$140$ MeV for $v_z = 0.2$, and $132$--$152$ MeV for
$v_z = 0.4$}. This is not longitudinal flow hardening the $p_T$ spectra, which
it cannot do, but a kinematic degeneracy: the longitudinal component of the
fluid four-velocity modifies the Lorentz-invariant factor $p^\mu u_\mu$ in the
distribution function, and the fit compensates by returning a higher $T$.
\rev{Temperature and transverse flow are strongly anti-correlated
($\rho_{T v_T}\approx-0.95$ at all energies), so they are jointly, not
independently, determined.}

\nw{A simultaneous fit of the $\pi^+$, $K^+$, and proton spectra with a shared
temperature and transverse flow sharpens the extraction and tests the kinetic
interpretation. The particle-mass ordering breaks the $T$--normalization
degeneracy of the single-species fit and tightens the parameter uncertainties
\fx{by factors of $3.7$--$5.3$ on $T$ and $2.2$--$2.9$ on $v_T$}, yielding a
lower and more physical kinetic freeze-out temperature
$T_{\rm kin}\simeq107$--$115$~MeV at $v_z=0$. \fx{These agree with STAR's
independently measured blast-wave temperatures at every energy: the differences
run from $+2.1$ to $-9.5$~MeV and remain within $0.9\sigma$ of the combined
uncertainties, STAR's own values carrying $\pm11$--$12$~MeV.} They sit well
below the STAR chemical freeze-out temperature, confirming that the present
analysis probes the kinetic rather than the chemical freeze-out surface.}

\rev{Throughout the scanned range $v_z \leq 0.4$ the extracted temperatures
remain at or below the QCD crossover temperature $T_c \approx 155$--$160$ MeV
estimated from lattice QCD, so that the hadron-resonance-gas description
underlying the model is internally consistent at every energy considered.
Scanning $v_z$ locates the critical value $v_z^{\,c}$ at which $T$ would reach
$T_c$ (Sec.~\ref{Section:corr}): $v_z^{\,c}\approx0.57$--$0.61$ at the lowest
energies, falling to \fx{$v_z^{\,c}\approx0.43$--$0.47$} at $39$ GeV in the
proton-only fit. The multi-species temperature is both lower and better
determined, and it is the corresponding bound $v_z\lesssim0.6$ that we adopt;
longitudinal velocities above it would drive kinetic freeze-out above the
crossover band, into a regime where the hadron-resonance-gas picture no longer
applies, and are on that basis physically disfavored.} \fx{A natural refinement
would impose $T \leq T_c$ as a hard constraint directly in the fit, yielding a
posterior on $v_z$ rather than the fixed-grid scan used here.}

\fx{Including the antiproton spectra with a normalization shared between $p$
and $\bar p$ removes the last external input: the $\bar p/p$ ratio fixes the
baryon chemical potential directly from the data, giving
$\mu_B \simeq 262$--$63$~MeV at kinetic freeze-out across the BES range. The
extracted $\mu_B/T$ agrees with STAR's chemical $\mu_B^{\rm ch}/T_{\rm ch}$ at
the $85$--$88\%$ level and is independent of $v_z$, exactly as expected if the
$\bar p/p$ ratio is frozen at the chemical surface and $\mu_B/T$ is conserved
during the subsequent hadronic cooling. Across all $v_z$ the freeze-out
trajectory in the $T$--$\mu_B$ plane follows the parametrization of
Cleymans~et~al.~\cite{PhysRevC.73.034905}, with $\mu_B$ decreasing
monotonically with $\sqrt{s_{NN}}$ and $T$ rising overall, subject to a mild
and statistically insignificant non-monotonicity near $11.5$~GeV
(Sec.~\ref{Section:Results}); it lies below both $T_c$ and the Cleymans
chemical-freeze-out line, as expected physically since kinetic freeze-out
occurs later and cooler than chemical freeze-out.

The resulting fully data-driven $(\rho_B,\varepsilon)$ trajectory, with
uncertainties propagated from the fit covariance, places kinetic decoupling at
the lowest BES energies in a dilute system, $\rho_B \simeq 0.03\,n_0$ (with
$n_0 = 0.16$~fm$^{-3}$), a factor of $\simeq25$ below the chemical-surface
compression maximum of Ref.~\cite{PhysRevC.74.047901}. The net baryon density
falls monotonically as $\sqrt{s_{NN}}$ rises while $\varepsilon$ remains nearly
energy-independent, placing the maximum of net-baryon compression at or below
the lower edge of the BES-I range, as found in
Ref.~\cite{PhysRevC.74.047901}. Collective flow raises these densities
substantially through the $T$--$v_z$ degeneracy: at $v_z = 0.4$, $\rho_B$ grows
by a factor of $\simeq4$ and $\varepsilon$ by $\simeq2.3$, the corresponding
$\rho_B$ enhancement in the hybrid evaluation---which pairs the fitted $T$ with
the chemical $\mu_B^{\rm ch}$---being $2.2$--$2.7$. The inferred thermodynamic
conditions therefore depend significantly on the assumed collective motion, and
the longitudinal flow must be constrained before the compression can be pinned
down precisely.

The integrated proton yields are essentially independent of $v_z$: going from
$v_z = 0$ to $v_z = 0.4$ raises the fitted temperature by about $11\%$ at every
energy yet changes the integrated yield by less than $0.02\%$, because the
normalization $\lambda$ absorbs the change almost exactly---the same
compensation that renders the fit insensitive to the input $\mu_B$.} Pion
production, by contrast, rises steadily with $\sqrt{s_{NN}}$, consistent with
the transition from baryon-dominated to meson-dominated particle production as
nuclear transparency increases with energy.

Several extensions of the present work are envisaged. In the near term, the
most pressing is the inclusion of additional hadron species, in particular
strange hadrons such as \fx{$K^-$}, $\Lambda$, and $\Xi$, in the simultaneous
fit. \fx{The required measurements already exist at exactly these five
energies: the $K^-$ spectra are published in the same data set used
here~\cite{PhysRevC.96.044904}, and $\Lambda$, $\bar\Lambda$, $\Xi^-$ and
$\bar\Xi^+$ spectra are available from STAR's strange-hadron
survey~\cite{PhysRevC.102.034909}.} This would allow the strangeness chemical
potential $\mu_S$ and the electric charge chemical potential $\mu_Q$, both set
to zero here, to be self-consistently determined\fx{, following the same
ratio-based strategy used for $\mu_B$ in Sec.~\ref{Section:mubextraction};
indeed Ref.~\cite{PhysRevC.102.034909} already extracts $\mu_S/T$ and
$\mu_B/T$ at freeze-out from antibaryon-to-baryon ratios of precisely these
species. Correcting the $p$ and $\bar p$ spectra for weak feed-down would
likewise remove the dominant known bias in the extracted $\mu_B$}. The latter
is expected to be non-negligible at lower energies, where baryon stopping and
the neutron-to-proton imbalance of the Au nucleus introduce an appreciable
isospin asymmetry. A global multi-species fit would also test the assumption of
a single common freeze-out surface, which may break down if different species
decouple at different times.

\fx{Both flow velocities enter the present description as single constants
rather than as profiles: $v_T$ is a single surface-like transverse velocity, as
the comparison with STAR's profile-averaged $\langle\beta\rangle$ in
Sec.~\ref{Section:multispecies} makes explicit, and $v_z$ is the magnitude of
the longitudinal flow velocity of the emitting elements. Only that magnitude
enters: because $\cosh$ is even and the acceptance window is symmetric about
$y=0$, the rapidity-integrated spectrum of Eq.~\eqref{Eq:eq15} is an exact even
function of $v_z$, so a source element moving forward and its backward-moving
mirror contribute identically and no forward--backward asymmetry is implied.
Extending the analysis to wider rapidity windows would accordingly be valuable,
though what such data would constrain is not the single magnitude $v_z$ but the
underlying \emph{distribution} of flow rapidities that it stands in for. Within
$|y|<0.1$ the two are indistinguishable, since $\cosh(y-y_k)$ may be replaced
by its value at $y=0$ throughout the window,precisely the origin of the
$T$--$v_z$ degeneracy above,whereas a boost-invariant continuum of flow
rapidities spanning $|y_k|\leq\mathrm{artanh}\,v_z$ predicts a markedly
different $p_T$ shape. Rapidity-differential spectra, together with net-baryon
rapidity distributions, would discriminate between these descriptions and so
replace the effective longitudinal parameter used here by a measured flow
profile.}

Finally, the present approach captures the freeze-out conditions through a
parametrized velocity field on a single surface element. \fx{The emission
itself is already of Cooper--Frye form, as noted below Eq.~\eqref{Eq:eq7}, and
would remain so in a more complete treatment: what a full $3+1$ dimensional
viscous hydrodynamic evolution with a baryon-density-dependent equation of
state (EoS) supplies is not a different hadronisation prescription but the
freeze-out hypersurface and the flow field upon it, determined dynamically by
the conservation laws rather than imposed, together with the shear and bulk
corrections $\delta f$ to the equilibrium distribution that the present
emission function omits. The parameters of Table~\ref{table:fitparams} would
then be replaced by their local counterparts $T(x)$, $\mu_B(x)$ and $u^\mu(x)$,
and the values quoted here read as emission-weighted averages over that
surface.}
\fx{In such a framework the EoS, and with it the adiabatic sound speed
$c_s^2 = (\partial P/\partial\varepsilon)_{s/n_B}$ at the finite net-baryon
densities relevant here, is an input rather than an output: it is supplied to
the evolution, which then propagates the pressure gradients that build the
collective flow. Its value would therefore be \emph{confronted with} the
measured spectra and freeze-out conditions rather than extracted from them.}

\fx{The ingredients for such a programme are already in place: lattice-matched
equations of state at finite net-baryon density~\cite{Monnai2019EoS}, families
of equations of state carrying a critical point in the three-dimensional Ising
universality class mapped onto them~\cite{Parotto2020EoS}, and $3+1$
dimensional viscous simulations with explicit baryon transport applied across
the BES energy range~\cite{Shen2018DynamicalIS}, assembled together with
critical fluctuations and freeze-out prescriptions into a common
framework~\cite{BEST2022Framework}. What remains open is not whether such
calculations can be carried out but where or whether the critical point
lies, since its location is not fixed by the lattice and enters as a parameter
of the constructed equation of state. The coordinates extracted here are
complementary in a specific sense: they carry no hydrodynamic input, so the
kinetic freeze-out surface of Table~\ref{table:mubfit}, and its $\simeq25$-fold
displacement below the chemical-surface compression maximum, are benchmarks
that such calculations should reproduce at their own decoupling surface
independently of the equation of state assumed at earlier times.}

\section{ACKNOWLEDGMENTS}
V.R. acknowledges financial support from Anusandhan National Research Foundation~(ANRF), India through Core Research Grant , CRG/2023/001309. \\

\appendix

\section{Covariant Statistical Mechanics and the Four-Temperature}
\label{App:covariant_stat_mech}

The covariant formulation of statistical mechanics was developed to resolve 
a fundamental ambiguity, how does temperature transform under a Lorentz 
boost, i.e., what is the temperature of a body in motion in Special 
Relativity? Two early proposals gave contradictory answers.

If one uses the ideal gas law, $PV = nRT$, to define temperature, and 
notes that volume contracts as $V = V_0/\gamma$ under a boost while the 
equation of state holds in every inertial frame, one obtains 
\begin{equation}
T(v) = \frac{T_0}{\gamma},
\label{Eq:Planck_T}
\end{equation}
where $T_0 \equiv T(0)$ is the rest-frame temperature and 
$\gamma = (1-v^2)^{-1/2}$. 

Alternatively, from the second law of thermodynamics, $dS = dQ/T$, the 
entropy $S = k_B \ln \omega$ counts microstates and is a Lorentz scalar, 
while $dQ$ transforms as the time-component of the heat four-vector, 
giving $dQ = \gamma\, dQ_0$. Demanding that $dS$ remain invariant then 
yields
\begin{equation}
T(v) = \gamma\, T_0.
\label{Eq:Ott_T}
\end{equation}

The resolution of this apparent paradox was addressed in a more systematic way by Israel and 
van Kampen using a fully covariant formulation by replacing the scalar 
inverse temperature $\beta = 1/T_0$ with the four-vector
\begin{equation}
\beta^\mu = \frac{u^\mu}{T_0},
\label{Eq:beta_four_vector}
\end{equation}
where $u^\mu = \gamma(1, \mathbf{v})$ is the fluid four-velocity satisfying 
$u^\mu u_\mu = 1$. The Gibbs ensemble for a relativistic gas is then 
constructed by maximizing the number of microstates $\omega$ subject to 
four-momentum conservation $\sum_j N_j p_j^\mu = P^\mu$ and particle 
number conservation $\sum_j N_j = N$. Introducing Lagrange multipliers 
$\alpha$ and $\beta_\mu$ for these constraints respectively, the 
stationarity condition yields
\begin{equation}
\ln N_j + \alpha + \beta_\mu p_j^\mu = 0,
\end{equation}
from which the Boltzmann factor follows immediately
\begin{equation}
N_j = \frac{N}{Z}\,e^{-\beta_\mu p_j^\mu}, \qquad 
Z = \sum_j e^{-\beta_\mu p_j^\mu}.
\end{equation}
The partition function $Z$ is a Lorentz scalar by construction, since 
$\beta_\mu p_j^\mu = p_j^\mu u_\mu / T_0$ is a scalar contraction. The 
Lorentz-invariant phase-space factor $p^\mu u_\mu$ appearing throughout 
the covariant fireball model (see Sec.~\ref{Section:dynamic_fireball}) arises directly 
from this structure with the identification~\eqref{Eq:beta_four_vector}.

\bibliographystyle{unsrt}
\bibliography{ref}

\end{document}